\documentstyle[aps,jpc,preprint]{revtex}

\begin{document}

	\title{Boundary Integral Methods for the
	Poisson Equation of Continuum Dielectric Solvation Models}


	\author{Lawrence R.  Pratt, Gregory J.  Tawa, Gerhard Hummer,
\break Angel E.  Garc\'{\i}a, and Steven A.  Corcelli}

	\address{Theoretical Division, Los Alamos National
Laboratory,Los Alamos, NM 87545}

	\author{LA-UR-95-2659}

	\date{\today}

	\maketitle

	\begin{abstract} This paper tests a dielectric model for
variation of hydration free energy with geometry of complex solutes in
water.  It works out some basic aspects of the theory of boundary
integral methods for these problems.  One aspect of the algorithmic
discussion lays the basis for multigrid methods of solution, methods
that are likely to be necessary for similarly accurate numerical
solution of these models for much larger solutes.  Other aspects of the
algorithmic work show how macroscopic surfaces such as solution
interfaces and membranes may be incorporated and also show how these
methods can be transferred directly to periodic boundary conditions.
This dielectric model is found to give interesting and helpful results
for the variation in solvation free energy with solute geometry.
However, it typically significantly over-stabilizes classic attractive
ion-pairing configurations.  On the basis of the examples and
algorithmic considerations, we make some observations about extension
of this continuum model incrementally to reintroduce molecular detail
of the solvation structure.  \end{abstract}



\section{Introduction}

This paper tests a dielectric model \cite{rashin-review,honig-review}
for variation of hydration free energy with geometry of complex solutes
in water.  It works out some basic aspects of the theory of boundary
integral methods for these problems and describes, demonstrates, and
tests such algorithms for solution of that macroscopic Poisson
equation.  It then makes some observations about extension of this
continuum model incrementally to reintroduce molecular detail of the
solvation structure.

Numerical solution of these models has been discussed many times before
\cite{miertus,warwicker,gilson_a,zauhar_a,klapper,gilson_b,%
pascual-ahuir,rashin_a,gilson_c,zauhar_b,davis,yoon-lenhoff,juffer,%
nicholls,snitzer,wang,oberoi,zhou,mdmp,you,simonson,tucker,bharadwaj}.
Dielectric continuum models are widely considered simplistic
descriptions of the solvation of charged or polar solutes in solutions.
Thus, it might not be obvious whether accurate solutions to the
governing Poisson equation are important.  However, when solved to
thermal accuracy these models can give helpful and interesting
descriptions of the variation of solvation free energy with geometry
\cite{phg,tp}.  These dielectric models are physical.  Furthermore,
their connection with molecular theory is known and straightforward
\cite{phg,tp,lbk,fbl,kw,hpga,hpgb}.  For these reasons the dielectric
models deserve to be exploited more thoroughly than they have been.
These interesting properties of the physical model require methods that
yield solvation free energies accurate on the thermal energy scale ($<
k_BT \approx 0.03 eV$ under the conditions of widest interest).
Experience has shown that this level of accuracy can be difficult to
achieve and requires careful numerical methods
\cite{mdmp,simonson,tucker,bharadwaj}.

The Poisson equation is a linear equation and the chief difficulty is
associated with the treatment of interface boundary conditions on a
molecular surface that can be irregular.  The possibility of
disconnection of the molecular surface with molecular rearrangement is
of utmost physical importance.  Viewed from a physical level, those
disconnections can produce microscopic solvent pockets that the model
treats as bulk continuum solvent.  It might be questioned whether the
model is significantly less accurate in those circumstances.  Viewed on
a technical level, those disconnections greatly complicate the
description of the molecular-solvent interface.  Because spatial
resolution in the description of that interface is important to
achieving accuracy in the solutions, it might be further questioned
whether it is feasible or desirable to obtain solutions of the models
that are sufficiently accurate to examine the effects of those
disconnections.  But boundary element approaches are expected to be
particularly appropriate for these purposes.  This has been clearly
expressed by Yoon and Lenhoff \cite{yoon-lenhoff}.  The development
here follows that idea but accomplishes the required boundary (or
surface) integrations utilizing two simple procedures:  (a) plaques are
defined on the basis of quasi-random number or `good lattice'
multidimensional integration methods
\cite{hammersley,keng,lubotzky,tichy,niederreiter,press}; and (b)
integrations over plaques are accomplished by Monte Carlo methods.
This results in the computational method that is simple and general,
that lends itself to demonstration of numerical convergence, and that
permits a systematic exploitation of coarse grained results.

The {\it a priori\/} theoretical support for these models is limited; it
has been mentioned above that they are physical and have a known simple
connection to molecular theory.  An additional point that is important
for molecular justification of the continuum theory is that the
interactions being treated, electrostatic interactions, are long-ranged
on the scale of molecular sizes \cite{rowlinson-widom,lsb}.  Because the
theoretical justification of these models is limited, empirical
validation for the anticipated applications is important.  Therefore,
this paper presents a number of examples of dielectric model predictions
of potentials of mean forces for rearrangements of molecular species and
reactive complexes in water.  It deserves emphasis that the issue of the
validity of these models for predictions of changes in solvation free
energy with changes in molecular geometry is distinct from those of
accuracy for one geometry or of accuracy for changes in charge
distribution at a fixed geometry.  Most of the previous applications,
electronic structure calculations and electron transfer theories, appear
to rely on answers to the latter question.  Preliminary results for
several of the examples below have appeared in our previous work
\cite{phg,tp}.  But with the exception of \cite{rashin89,ford} this is
the only available work that obtains and studies potentials of mean
forces, {\it i.e.\/} variations of solvation free energies with
geometry, accurate on the thermal energy scale.  Note also \cite{thg}.
Reference \cite{rashin89} provided an important motivation for the
present work.

\section{Physical Model}

We consider a molecule in a solution environment and
identify that molecule as the `solute' of interest.  A solute volume is
defined on the basis of its geometry.  Partial charges describing the
solute electric charge distribution are positioned within this volume.
The Poisson equation of interest is
\begin{equation} \nabla \bullet \varepsilon ({\bf r})\nabla \Phi ({\bf
r})=-\;4\pi \rho({\bf r}) \label{poisson} \end{equation}
where $\rho ({\bf r})$ is the density of electric charge associated
with the solute mole\-cule, the function $\varepsilon({\bf r})$ gives
the local value of the dielectric constant, and the solution $\Phi({\bf
r})$ is the electric potential.

\subsection{Integral equations for the model}

When contributions from an external surface vanish, the fundamental
solutions of this Poisson equation are solutions of the integral
equations:  \begin{mathletters} \label{inteq} \begin{equation}
\varepsilon({\bf r})G({\bf r},{\bf r}')= G^{(0)}({\bf r},{\bf r}')
+\int\limits_V { {\nabla ''G^{(0)}({\bf r},{\bf r}'')} \bullet \left[
{{{\nabla ''\varepsilon ({\bf r}'')} \over {4\pi \varepsilon ({\bf
r}'')}}} \right] {\varepsilon ({\bf r}'')G({\bf r}'',{\bf r}')} d^3r''}
, \label{inteqa} \end{equation} \begin{equation} G({\bf r},{\bf
r}')=G^{(0)}({\bf r},{\bf r}') -\int\limits_V {\left[ {\nabla
''G^{(0)}({\bf r},{\bf r}'')} \right]\bullet \left( {{{\varepsilon ({\bf
r}'')-1} \over {4\pi }}}\right) \left[ {\nabla ''G({\bf r}'',{\bf r}')}
\right]d^3r''} , \label{inteqb} \end{equation} \begin{equation} G({\bf
r},{\bf r}') \varepsilon ({\bf r}')=G^{(0)}({\bf r},{\bf r}')
+\int\limits_V G^{(0)}({\bf r},{\bf r}'') \left[ {\nabla ''\varepsilon
({\bf r}'')\over 4\pi\varepsilon ({\bf r}'')} \right] \bullet {\nabla
''G({\bf r}'',{\bf r}')\varepsilon ({\bf r}') d^3r''}.  \label{inteqc}
\end{equation} \end{mathletters} The derivation of these equations is
discussed in an appendix.  Here $G({\bf r},{\bf r}')$ is the Green
function for Eq.\ (\ref{poisson}) --- the electric potential at ${\bf
r}$ due to a unit source at ${\bf r}'$ --- and $G^{(0)}({\bf r},{\bf
r}')$ is the Green function for the case where $\varepsilon({\bf r}) =
1$.  These equations apply both to the `unbounded case' and to periodic
boundary conditions.  For the latter case the potential is only required
in a cell of volume V and the integral need only cover that volume.  But
in the latter case the electric potential due to a unit source
``$G^{(0)}({\bf r},{\bf r}')$'' is then the traditional Ewald potential.
That result also is discussed in an appendix.  It will be worthwhile to
note that Eqs.\ (\ref{inteq}) are specializations of the more general
relations that arise from considering how to calculate $G({\bf r}, {\bf
r}')$ for a specified $\varepsilon({\bf r})$ with the help of a
separately known $G^{(0)}({\bf r}, {\bf r}')$ corresponding to
$\varepsilon^{(0)}({\bf r})$.  Those more general relations are:
\begin{mathletters} \label{refinteq} \begin{eqnarray} \varepsilon ({\bf
r}) G({\bf r},{\bf r}') & = & \varepsilon^{(0)}({\bf r}) G^{(0)}({\bf
r},{\bf r}') \nonumber \\ & + & \int\limits_V \varepsilon^{(0)}({\bf r})
\nabla '' G^{(0)}({\bf r},{\bf r}'') \bullet ({1 \over 4 \pi }) \bigl[
\nabla'' \ln \left( {\varepsilon ({\bf r}'') \over \varepsilon^{(0)}
({\bf r}'')} \right) \bigr] \varepsilon({\bf r}'') G({\bf r}'',{\bf r}')
d^3r'' , \label{refinteqa)} \end{eqnarray} \begin{equation} G({\bf
r},{\bf r}')=G^{(0)}({\bf r},{\bf r}') -\int\limits_V {\left[ {\nabla
''G^{(0)}({\bf r},{\bf r}'')} \right]\bullet \left( {{{\varepsilon ({\bf
r}'')- \varepsilon^{(0)}({\bf r}'')} \over {4\pi }}}\right) {\nabla
''G({\bf r}'',{\bf r}')} d^3r''} , \label{refinteqb} \end{equation}
\begin{eqnarray} G({\bf r},{\bf r}') \varepsilon({\bf r}') & = &
G^{(0)}({\bf r},{\bf r}')\varepsilon^{(0)}({\bf r}') \nonumber \\ & + &
\int\limits_V G^{(0)}({\bf r},{\bf r}'') \varepsilon^{(0)}({\bf r}'')
({1 \over 4 \pi}) \nabla'' \ln \left( {\varepsilon ({\bf r}'') \over
\varepsilon^{(0)} ({\bf r}'')} \right) \bullet \nabla '' G({\bf
r}'',{\bf r}') \varepsilon({\bf r}')d^3r'' .  \label{refinteqc}
\end{eqnarray} \end{mathletters} Again, the derivation of these
equations is discussed in an appendix.  These equations will be of
subsequent help in the design of numerical methods to treat systems for
which the solute volume is not efficiently described as a finite
collection of finite spheres, {\it e.g.\/}, systems that include slabs
and cylinders.

\subsection{Molecular Volume and Boundary}

In the customary applications of dielectric models a molecular volume is
defined and the region outside this molecular volume is assigned a local
value of the dielectric constant, $\varepsilon({\bf r})$, equal to the
measured dielectric constant of the solution.  It is with the definition
of the molecular volume that molecular scale information of the
solvation structure, otherwise external to the model, is incorporated.
It should be recognized that the defined molecular volume typically will
change with the thermodynamic state of the system, with the structure of
the solute, with solute-solvent interactions, and with correlations
among solvent molecules.  We comment later on the statistical
thermodynamical underpinning of the model that clarifies the molecular
origin of that information.  The description of the molecular surface
should be simple and should permit disconnection with molecular
rearrangement.  Smoothness is desirable.  The most popular
previous choice has been the Connolly surface \cite{connolly} defined as
a surface of contact between a spherical probe and the van der Waals
surface of the molecule.  We have chosen instead to define the volume of
a molecular complex in terms of a collection of spheres, each member $s$
of that collection having an individually specified radius, $R(s)$, and
center, ${\bf c}(s)$.  The volume enclosed by spheres in this collection
is the `molecular volume.'  It is expected that the molecular volume
will contain the source charges but that limitation could be
straight\-forwardly relaxed.  The molecular surface is defined as the
boundary of the molecular volume.  Generally, some of the surfaces of
the individual spheres will be inside other spheres; some spherical
surface will be buried and will not contribute to the molecular surface.

Our considerations for making this choice of the molecular surface
rather than the Connolly surface are that it is simpler and no less
physical.  The Connolly surface has the advantage of better smoothness
for compact molecular conformations.  But it has the disadvantage that
disconnection upon continuous conformational change either leaves sharp
`cusps' or, if those cusps are arbitrarily shaved-off, the volume
changes discontinuously.  Since the statistical thermodynamic connection
of the defined molecular surface to the solvation structure is somewhat
indirect, involving solvent molecular sizes also, the differences
between the present choice of molecular surface and the Connolly surface
is not significant physically.

This choice, and the Connolly choice also, yields a molecular surface
that is generally not a Liapunov surface \cite{jaswon}.  That smoothness
condition is an important feature of general abstract discussions of
boundary element methods \cite{jaswon}.  The importance of the
smoothness of the surface for general discussions can be exemplified by
considering the case of the traditional Dirichlet problem:  solution of
$\nabla^2 \varphi({\bf r})=0$ within a closed volume with $\varphi({\bf
r})$ specified on the boundary.  If, for example, the boundary does not
have a unique surface normal then discussions involving the normal
derivative of $\varphi({\bf r})$ on the boundary become vexed.  The
significance of this point for our applications can be discussed on the
basis of Eq.\ (\ref{inteqa}).  The homogeneous contribution
$G^{(0)}({\bf r}, {\bf r}')$ is continuous, smooth, and differentiable
away from sources and, on the assumption that no sources are located
precisely on the surface, on the boundary of the molecular volume also.
In order to isolate potential problems we can focus on the integral of
Eq.\ (\ref{inteqa}).  We notice that the discontinuous function
$\varepsilon({\bf r})$ appearing in the denominator can be cancelled
exactly and that the factor $G({\bf r}'', {\bf r}')$ is expected on the
basis of Eq.\ (\ref{poisson}) to be continuous and finite away from
sources and thus on the boundary in our problem.  [In fact, we can
typically arrange it to be non-negative.  Then Eq.~(\ref{inteqa}) can be
a natural basis of Monte Carlo calculations.]  If we agree not to
require computation of the electric field precisely at points where no
unique surface normal exists, then the only potential difficulty to
consider is associated with the integrability of the surface dipole
contribution to the electric potential at a point off the boundary at
positions of the sources.  But those contributions are clearly
integrable and thus only a problem of practical difficulty.

\subsection{Comments}

Some points of note about Eqs.\ (\ref{inteq}) are:  first, Eq.\
(\ref{inteqb}) is simple to interpret physically.  It says that the net
electrostatic potential of a charge is composed of the potential due to
polarization induced in the medium by electric fields there in addition
to a direct, or bare interaction.  Second, for the conventional
applications $\varepsilon({\bf r})$ is a sharp step at the molecular
surface and thus the integrals of Eqs.\ (\ref{inteqa}) and
(\ref{inteqc}) are surface integrals over the molecular surface.
Although these equations are obviously twins, they offer different
physical pictures of the problem.  The integral of Eq.\ (\ref{inteqa})
describes a discontinuous function of ${\bf r}$; that discontinuity is
consistent with the obvious discontinuity of the left-side of Eq.\
(\ref{inteqa}).  The integrand is a continuous function, however, for
{\bf r} not on the boundary.  For {\bf r} inside the molecular volume
where $\varepsilon({\bf r})=1$, the integrand of Eq.\ (\ref{inteqa}) can
be viewed as the potential due to a dipolar surface density $G({\bf
r},{\bf r}')$.  The integrand of Eq.\ (\ref{inteqc}) can be considered
to be the potential due a surface charge density $\bigl( 4\pi\varepsilon
({\bf r}'') \bigr) ^{-1} \nabla ''\varepsilon ({\bf r}'') \bullet \nabla
''G({\bf r}'',{\bf r}')\varepsilon ({\bf r}')$.  For this case, the
integrand varies discontinuously with normal displacement through the
molecular surface because the normal electric field is discontinuous at
a sharp step in $\varepsilon({\bf r})$.  If we evaluate that negative
gradient normal to the molecular surface by assuming that it is half the
sum of the inside and outside values, we obtain from Eq.\ (\ref{inteqc})
the equation that provides the basis for the most widely used of
previous boundary element calculations
\cite{rashin-review,miertus,zauhar_a,rashin_a,zauhar_b,yoon-lenhoff,%
juffer,wang}.  This is discussed in an appendix.  Eq.~(\ref{inteqc}) can
be written more physically for the electric potential of
Eq.~(\ref{poisson}) as \begin{equation} \Phi ({\bf r})=\Phi ^{(0)}({\bf
r})+\int\limits_V {G^{(0)}({\bf r},{\bf r}') \left( {{{\nabla
'\varepsilon ({\bf r}')} \over {4\pi \varepsilon ({\bf r}')}}} \right)
\bullet \nabla '\Phi ({\bf r}')d^3r'} \label{5} \end{equation} with the
requirement that $\varepsilon({\bf r})=1$ wherever $\rho_f({\bf r}) \ne
0.$ Here $\Phi ^{(0)}$ is the electric potential in the absense of the
medium.

	In constructing and analyzing practical solutions is it helpful
to recognize simple applications of Gauss's law, for example to
Eq.~(\ref{inteqa}) \begin{equation} \int\limits_V {\nabla
'G^{(0)}(r,r')\bullet \nabla '\varepsilon (r')d^3r'}=-4 \pi \Delta
\varepsilon\eta({\bf r}) \label{glaw} \end{equation} where $\eta({\bf
r}) $ is one or zero if the observation point $ ({\bf r}) $ is inside or
outside, respectively, the molecular volume.

\section{Examples}

All examples here used $\varepsilon = 77.4$ in the solvent.  The
quantities presented in these examples are the potentials of the mean
forces obtained as \begin{equation} W = U + ({1 \over 2}) \int {\rho (
{\bf r} ) \bigl( \Phi( {\bf r} ) - \Phi^{(0)}( {\bf r} ) \bigr) d^3{\bf
r} }.  \label{pomf} \end{equation} U is the static energy of the solute
in the absence of the solvent.  Typically, a constant contribution to
$W$ is adjusted so that this potential of mean force is zero at some
asymptotic separation of interest.  The distribution of free charge
$\rho ( {\bf r} ) $ is a sum of partial charges so the integral of Eq.
(\ref{pomf}) reduces to a sum over those charges.  That is not an
important limitation, clearly.  Except where noted, we used the same
charge distribution employed in obtaining the simulation results used as
a comparison.

	All our results below were obtained on the basis of either
Eqs.~(\ref{inteqa}), or (\ref{inteqc}), or both, utilizing methods
described subsequently.  Wherever possible we used the radii-parameters
of Ref.~\cite{rashin85}.  Otherwise we adopted van der Waals radii
arbitrarily and will note those values in the specification of the
example calculations.  We note that knowledge of the helpful work
Ref.~\cite{lim} would probably have permitted better initial estimates
of the required radii-parameters.  However, we reiterate our expectation
that variation of hydration free energies with geometry place a
different demand on the dielectric model and upon radii-parameters than
do total free energies or total free energy changes upon reaction.  We
expect that qualitative conclusions drawn below would not be changed by
reasonable alterations of the radii-parameters used.

	Energy units on the following graphs are either k$_B$T or
kcal/mol.  That choice was governed by the energy units chosen in
reporting the original results that we are using for comparison.  For
all the results here k$_B$T $\approx 0.59$kcal/mol and that smaller unit
was typically adopted where more refined analyses were pursued.

	We note that practioners are not yet unanimous regarding the
most appropriate boundary conditions to use for electrostatic
interactions in simulation calculations of the sort we rely on here.
Different boundary condition choices can lead to quite different
results.  We have exercised some judgement in choosing simulation
results to use.  In some cases, we have not used simulation results that
are substantially the same as those we have used.  In some other cases,
however, we have excluded simulation results that display obvious
dubious features, even though those features do not presently have a
clear explanation.

	Our results below are based upon solutions of
Eq.~(\ref{poisson}) that are believed to be accurate to about 0.3 k$_B$T
for the total solvation free energy for the worst cases.  The principal
difficulty we have encountered is the following:  Numerical solution of
Eq.~(\ref{poisson}) requires replacement of a continuous problem with a
discrete one.  The results of different calculations can change
practically discretely \cite{simonson,tucker,bharadwaj} unless special
care is taken to avoid that and this is especially important for changes
in geometry.  For example, the position or number of lattice points can
change discontinuously depending on the geometry and how the numerical
problem is organized.  Such changes can be small on the scale of the
total solvation free energy, {\it e.g.\/} 0.5\% of the whole, but large
and puzzling on the scale of the potentials of mean force.  None of
those features are evident in our results below.  Thus, we believe our
results presented here are the correct solution of the dielectric model
for the problems considered.  A major emphasis of the subsequent
discussion of numerical methods is placed upon getting correct and
correctly continuous variation of the free energy with geometry.

\subsection{Na$^+ \cdots$ Cl$^-$ pair potential of mean force in water}

A variety of theoretical results
\cite{phg,tp,rashin89,hummer92,guardia91a,guardia91b,pr} for the
potential of mean force between a Na$^+ \cdots $ Cl$^-$ ion pair in
water are shown in Fig.\ \ref{nacl.fig}.  We note that the XRISM
\cite{pr} and the dielectric model calculations agree in properly
describing the asymptotic $1/\varepsilon r$ behavior.  All results show
a contact minimum at about 2.6-2.8 \AA.  The repulsive, short-distance
side of this contact minimum is due to the effect of the solvent in
pulling the ion pair apart.  The direct ion core overlap repulsion
begins to be significant only for distances less than 2.5 \AA.

	The two dielectric model calculations used different molecular
surfaces.  The earlier work of Rashin \cite{rashin89} employed a
Connolly surface \cite{connolly}.  The later calculation \cite{phg} used
the van der Waals surface.  The comparison of Fig.  \ref{nacl.fig} shows
that those differences in the molecular surfaces are important in
the region of the free energy barrier.  However, the close equality of
those results near the contact minimum suggests that the differences in
the molecular surfaces used are not important in that region of
inter-ionic separations.

On this basis, we restrict our conclusions here to the region of the
contact minimum.  We conclude that the dielectric models predict that
contact-paired ions are too stable relative to the dissociation limit.
The XRISM approximation is somewhat better in this regard but still errs
in predicting too much stability for the contact ion pair.  These errors
are appreciable on the $kT$ energy scale so that predictions of
probability densities would be in error by factors of about 3.

	We note that including a modeled polarizability of the solvent
and solute species does not seem to have a big effect on the predicted
Na$^+ \cdots $ Cl$^-$ pmf for distances less than about 8\AA\
\cite{smith:94}.

\subsection{Cl$^- \cdots$ Cl$^-$ pair potential of mean force in water}

A variety of theoretical results \cite{rashin89,hummer92,guardia91b,pr}
of the potential of mean force between a Cl$^+ \cdots $ Cl$^-$ ion pair
in water are shown in Fig.\ \ref{clcl.fig}.  All results except
`molecular dynamics 92' agree in properly describing the asymptotic
$1/\varepsilon r$ behavior.  Those exceptional molecular dynamics
results were obtained at a non-zero salt concentration.

The two dielectric model calculations used different molecular surfaces:
Ref.  \cite{rashin89} adopted a Connolly surface \cite{connolly}; the
present adopted the van der Waals surface with the same
radii-parameters.  The dielectric model predictions are entirely
repulsive.

\subsection{t-butyl$^+ \cdots$ Cl$^-$ potential of mean force in water}

Fig.~\ref{tbut.fig} shows theoretical results for approach of a
Cl$^-$ ion to a planar t-butyl$^+$ for ion-pairing configurations.  The
Monte Carlo and XRISM results are taken from Ref.~\cite{wljtb}.  For the
dielectric model calculations, the radius for the tertiary carbon was
taken as $R = 1.85$\AA; for the united atom methyl groups those radii
were $R = 2.00$\AA\ \cite{amber}; for the chloride the Rashin-Honig
\cite{rashin85} value $R = 1.937$\AA\ was used.  Again, the dielectric
model predicts too much stability of contact pairs relative to the
dissociated fragments.  The XRISM approximation is somewhat better but
errs in the same direction by about 4k$_B$T or about a factor of 55 in
the associated radial distribution function.

	The results of Ref.~\cite[Figure 5]{ford} cannot be conveniently
placed on this graph.

\subsection{Na$^+ \cdots$ dimethylphosphate$^-$ ion pair potentials of
mean force in water}

For the Na$^+ \cdots$ dimethylphosphate$^-$ complex ion pair two
sets of theoretical results are available \cite{nadmpa,nadmpb}.  For
both cases, the conformation of the dimethylphosphate anion was held
rigidly in the {\it gg\/} structure of the fragment as described by
electronic structure calculations \cite{nadmpc}.  In the first
`asymmetric' case the Na$^+$ ion was brought up along a P-O bond of an
exposed oxygen atom; see Fig.~\ref{nadmpasm.fig}.  In the second
`symmetric' case the Na$^+$ ion was brought up along the bisector of the
O-P-O angle involving exposed oxygen atoms; see Fig.~\ref{nadmpsym.fig}.
Beyond the sodium radius \cite{rashin85} of $R = 1.68$\AA, the radii
used in this case were van der Waals radii of Ref.~\cite{amber}:
phophorus, $R = 2.10$\AA; oxygen(ester), $R = 1.65$\AA;
oxygen(phosphate), $R = 1.60$\AA; united atom methyl, $R = 2.00$\AA.  In
the asymmetric case, the {\it gaussian bath\/} approximate theory agrees
closely with the molecular dynamics results.  The XRISM and dielectric
model over-stabilize the asymmetric contact ion pairs.  For the
symmetric approach the molecular dynamics results are different from all
the other theoretical results.  Those theoretical results err by about
4k$_B$T or about a factor of 55 in the first peak for the associated
radial distribution function.

	We have not used the results of Ref.~\cite{nadmpd}.

\subsection{Chloride exchange in methyl chloride by symmetric S$_N$2 in
water}

Fig.  \ref{methchl.fig} shows available results for the model
S$_N$2 reaction by a symmetric exchange of chloride in methylchloride
along a linear reaction path \cite{sn2a,sn2b,sn2c}.  Beyond the chloride
radius $R = 1.937$\AA \cite{rashin85}, we used radii \cite{amber}:
hydrogen, $R = 1.00$\AA; carbon, $R = 1.85$\AA.  We used the partial
charges of \cite{sn2a} except for reaction coordinates in the range
-1\AA $<r<$ 1\AA\ where used the revised model charges of \cite{sn2c}.
Our dielectric model result appears to be in good agreement with the
earlier work of Ref.~\cite{ford} that was based upon AM1 electronic
structure model coupled to the dielectric medium.  The agreement seen
between dielectric models and the Monte Carlo results is good, better
than that agreement between XRISM and Monte Carlo.

\subsection{Nucleophilic addition of hydroxide to formaldehyde in water}

Fig.  \ref{formaldehyde.fig} shows available results for the
attack of formaldehyde by hydroxide through a S$_N$2 mechanism
\cite{forma,formb}.  We used radii \cite{amber}:  hydrogen, $R =
1.00$\AA; oxygen (hydroxide), $R = 1.65$\AA; oxygen (carbonyl), $R =
1.60$\AA; carbon, $R = 1.85$\AA.  The prediction of the dielectric model
is qualitatively correct but with large quantitative errors.  It has
been shown that these quantitative errors can be empirically corrected
by a reparameterization that pays no attention to physicality\cite{tp}.
The XRISM results are quantitative in bettter agreement with the Monte
Carlo data but that XRISM approximation still incurs substantial
quantitative errors, over-stabilizing the compact configuration relative
to the dissociated species.

\subsection{Conformational equilibrium of N-methylacetamide in water}

All the examples here have been considered in order to test the
theoretical model and not as a study of the particular system.  Thus,
the fidelity of the interaction model used for the molecular simulations
to laboratory experiments has been less a concern than the correctness
of the simulation results for the interaction model used.  This point
deserves particular emphasis for the N-methylacetamide example
considered next.  The conformations and interaction models that have
been studied in this case appear not to be very realistic.  However, the
value of the simulation work as a benchmark result against which
theoretical models can be compared remains.

What has been done is to define two planes as follows:  plane \#
1 is defined as the plane of the H(N), CH$_3$(N), and the C(carbonyl).
Plane \# 2 is the plane of the CH$_3$(C), O, and the N.  Then with the
C(carbonyl) and the N atom fixed, all other bond lengths and angles
fixed, and all force field parameters such as charges fixed, these
planes are rotated relative to each other by an angle $\omega$ about the
C(carbonyl)-N line.  The {\it trans\/} conformation, which is the lowest
energy conformation in the gas phase, corresponds to $\omega =
180^\circ$.  The {\it cis\/} conformation is $\omega = 0^\circ$.

The infelicities of this study as a description of laboratory
experiments are essentially two.  The first was discussed in
Ref.~\cite{nmaaa}.  The description followed here using partial charges
that are independent of conformation predicts that the {\it trans\/}
isomer has a larger dipole moment than the {\it cis\/} isomer.
Electronic structure calculations order the sizes these dipole
moments oppositely \cite{nmaaa,nmaab}.  Because the modelled dipole
moments do not display a correct dependence on conformation, the implied
solvation contributions appear to be incorrect in preferentially
stabilizing the {\it trans\/} conformation.  When this feature of the
molecular model is revised, the solvated free energies of the {\it
cis\/} and {\it trans\/} conformers are substantially the same
\cite{nmaaa}.

The second difficulty has to do the with path chosen in these
studies for interconversion of the two conformers of interest.  A more
detailed study would show that a natural path for interconversion of
these isomers would be very different from the simple rotation of the
planes defined above.  Thus the results here should not be taken as a
realistic description either of the net free energy difference between
the two isomers or of the variation of free energy along a natural
reaction path.

However, a most important attribute of a test of theoretical
model is that it is applied for precisely the same circumstances that
characterize the experimental data.  The available data are
satisfactory for that purpose and such a comparison is given in
Fig.~\ref{fig.nmaa} \cite{nmaac}.  The radii-parameters used were
$R_{H(N)} = 1.25$\AA, $R_{acetyl-methyl} = 1.955$\AA, $R_{N-methyl} =
1.90$\AA, $R_{C} = 1.875$\AA, $R_{N} = 1.625$\AA, $R_{O} = 1.48$\AA.
The comparison shows that the dielectric model and all the other
theories shown, except the XRISM approximation, give the right net
solvation free energy.  The apparent structural feature in the region
$90^\circ < \omega < 120^\circ$ is missed by all the theories.

\section{Methods}

We specialize Eqs.\ (\ref{inteqa}) and (\ref{inteqc}) or (\ref{5}) by
taking the observation point {\bf r} definitely just inside or outside
that molecular surface defined above.  In each case we then obtain a
closed integral equation that can be solved to provide the information
that determines the desired solution everywhere else.  The integral in
those boundary integral equations goes over the two dimensional surface
of the molecular complex.  The surface integral might be approximated by
sampling the molecular surface using pseudo-random number series,
quasi-random number series, or lattice rules
\cite{hammersley,keng,lubotzky,tichy,niederreiter,press}.  These
sampling points are equidistributed on the surface of the molecular
volume.  The development below eventually advocates using one of the
latter two methods to define plaques on the molecular surface and then
the use of Monte Carlo methods --- with pseudo-random number series ---
to accomplish integrations over the plaques.  For clarity of exposition,
however, we first describe the direct `sampling' of the integrand.  This
sampling permits the evaluation of integrals over the molecular surface.
In any of the cases anticipated, the integrand linearly involves the
unknown quantities such as $G({\bf r}, {\bf r}')$.  Therefore, these
methods of approximating the integral produce a finite set of linear
equations that can be solved by standard methods.

\subsection{Direct Sampling of the Molecular Surface} We begin by
writing out the equations used with direct, or elemental, sampling of
the molecular surface, treating Eq.\ (\ref{inteqa}) first.  We introduce
the notation \begin{equation} \int_V G^{(0)}({\bf r }, {\bf r}')
\rho({\bf r}') d^3 r' \equiv \Phi^{(0)}( {\bf r}), \label{phi-zero}
\end{equation} and similarly for $\Phi({\bf r})$.  We here focus on
${\bf r}$ infinitismally outside the molecular surface.  We will denote
the external (solution) dielectric constant by $\varepsilon_s$ and the
molecular volume will be assigned a dielectric constant $\varepsilon_m$.
An appropriate discretized version of Eq.\ (\ref{inteqa}) is then
\begin{equation} \varepsilon_s \Phi({\bf r}_i) = \Phi^{(0)}({\bf r}_i) +
\sum_j w_{ij}\Phi({\bf r}_j) \label{p-phidis} \end{equation} where the
coefficients are obtained as \begin{equation} w_{ij} = { ( {\bf r}_i -
{\bf r}_j ) \bullet {\hat {\bf n} }_j \over |{\bf r}_i - {\bf r}_j |^3 }
\times { (\varepsilon_s - \varepsilon_m) \over 4 \pi } \times \left(
{4\pi R(s_j)^2 \eta({\bf r}_j)\over M(s_j)} \right) , \qquad i \ne j .
\label{p-odmat} \end{equation} Here $R(s_j)$ is the radius of the sphere
$s_j$ whose surface is sampled by the $jth$ sampling point and ${\hat
{\bf n} }_j$ is the unit vector normal to the molecular surface at that
sampled point.  $\eta({\bf r}_j)$ is an indicator function for the
non-buried surface; it is {\it zero\/} if the $jth$ sampling point is on
buried surface and {\it one\/} if it is not.  $M(s_j)$ is the number of
sampling points used for sphere $s_j$.  The left-most factor of
$\varepsilon({\bf r})$ of Eq.\ (\ref{inteqa}) is $\varepsilon_s$ because
we are utilizing this equation for ${\bf r}$ just outside the molecular
surface.  The diagonal terms in this matrix of coefficients are simply
obtained by noting that the potential just outside a uniform sheet of
dipoles is $2 \pi \times$ (the surface density of moment).  Thus we have
\begin{equation} w_{jj} = 2 \pi \times {(\varepsilon_s - \varepsilon_m )
\over 4 \pi} \times \eta({\bf r}_j) = (\varepsilon_s - \varepsilon_m
)\eta({\bf r}_j)/2 .  \label{p-dmat} \end{equation} With these
specifications, Eq.\ (\ref{p-phidis}) can be solved directly to yield
the potential on the molecular surface.  It is clear that sampling
points on buried surface do not contribute to the physical resolution of
the potential at the molecular surface beyond entering into evaluation
of the matrix elements through the normalization numbers $M(s_j)$.  As
long as this normalization is properly maintained, sampling points on
buried surfaces can be preemptively deleted from the list of sampling
points.

Note how this procedure works when applied to the simplest problem:  the
molecule is a single sphere with a charge at the center; $\varepsilon_s
= \varepsilon$ and $\varepsilon_m = 1 $.  We know that the potential is
the same everywhere on the molecular surface.  If we sample the surface
only at only one point ($M(s_j) = 1$) we find that the potential at the
sampled point is $$ {1 \over R} \times {1 \over \varepsilon - w_{11}} =
{2 \over (\varepsilon+1)R}.  $$ The correct answer is $1/\varepsilon R$.
This example suggests the slight modification of Eq.\ (\ref{p-dmat})
that arises from a slightly more refined estimate of the solid angle
subtended by the molecular surface element at the {\it jth\/} sampling
point:  \begin{equation} w_{jj} = (\varepsilon_s - \varepsilon_m )
(1-M(s_j)^{-1/2}) \eta({\bf r}_j)/2 .  \label{ref-dmat} \end{equation}
This refinement can be derived by assuming that the $jth$ plaque has
area $4\pi R(s_j)^2/ M(s_j)$ and that area is bounded by a circle, {\it
i.e.\/} each plaque is a circular cap with the spatial extent determined
by the average.  With this modification and only one sampled point
($M(s_j)=1$), the present method produces the exact answer for the
potential at the surface when applied to this simplest problem.  With
this modification, the simple `one-point' calculation satisfies the row
sum rule $\Sigma_{k} w_{jk}=0$ that follows from the application of
Gauss's law Eq.~(\ref{glaw}).

This method works straightforwardly.  The continuum model results of
Ref.  \cite{phg} show what can be achieved with this primitive approach.
We note that the difference between Eq.\ (\ref{p-dmat}) and
(\ref{ref-dmat}) was observed to be important in those calculations.
One limitation is that to get accurate solutions the matrix $w_{ij}$
sometimes must be large, {\it e.g.\/} 10$^4 \times$10$^4$.  For our
present interests, `accurate' has been gauged by the results for the
solvation free energy.  For the case $\varepsilon_m=1$ this is
\begin{equation} \Delta \mu =\left( {{1 \over 2}} \right)\int\limits_V
{\rho ({\bf r})\left( {\Phi ({\bf r})-\Phi^{(0)} ({\bf r})}
\right)d^3r}, \label{dmu} \end{equation} should be correct on the
thermal energy scale $k_B T$ and should display reasonable variations on
this energy scale with variations of the solute geometry.  Other
properties of the solutions of these equations can be more sensitive.
The electric fields in the interior of the solute are examples of such
properties.  We close this section by considering another of those more
sensitive properties but still a property of physical interest, a
molecular polarizability modelled on dielectric idealizations.

The method above offers a simple way to model the polarizability of a
solute molecule, by assigning a dielectric constant $\varepsilon_m \ne
1$ to the molecular volume \cite{sharp}.  This possibility is helpful
for the physical fidelity of these models but also provides a tougher
test of the numerical methods used.  To establish a value of
$\varepsilon_m$ on the basis of a prescribed information about the net
polarizability $\alpha$ we proceed as follows.  Choose $\varepsilon_s=1$
and augment $\Phi^{(0)}({\bf r})$ to include a uniform external electric
field:  \begin{equation} \Phi^{(0)}({\bf r}) \leftarrow \Phi^{(0)}({\bf
r}) - {\bf r} \bullet {\bf E}^{(0)} .  \label{augment} \end{equation} We
then use the methods here to determine $\Phi({\bf r})$ far from the
solute molecule.  More specifically, we calculate \begin{equation}
{{\partial p_\gamma } \over {\partial E_\eta ^{(0)}}}\equiv \alpha
_{\gamma \eta } \label{def-pol} \end{equation} where the dipole moment
${\bf p}$ is computed from a specified origin.  We note that the
equation to be solved, consider Eq.\ (\ref{inteqa}) for example, is
linear.  Therefore a {\it response\/} to a particular contribution
$\Phi^{(0)}({\bf r})$ can be computed independently of other
contributions.  Thus, we can calculate $\alpha_{\gamma \eta}$ without
consideration of the charge distribution of the solute.  It is a
property only of the parameter $\varepsilon_m$ and the shape of the
molecular volume.  It is independent of a coordinate system origin.
With the specifications here we solve \begin{equation} \varepsilon_m
\Phi_\eta({\bf r}_i) = - {\bf r}_i \bullet \hat {\bf e}_\eta + \sum_j
w_{ij}\Phi_\eta({\bf r}_j) \label{leq-pol} \end{equation} as above.
$\hat {\bf e}_\eta$ is a unit vector along axis $\eta$.  We then form
the desired modeled net polarizability from \begin{equation} \alpha
_{\gamma \eta }=\left( {1-\varepsilon _m} \right)\hat {\bf
e}_\gamma\bullet \sum\limits_i \left( {{\bf r}_i-{\bf c}(s_i)}
\right){R(s_i) \over M(s_i)}\Phi _\eta ({\bf r}_i).  \label{con-pol}
\end{equation} The distance of the plaque point ${\bf r}_i$ from the
center of its sphere ${\bf c}(s_i)$ arises because the surface dipoles
are oriented along the outward pointing normal to the surface.  As an
example, consider a sphere of radius R, sample the surface at only one
point ${\bf r}$ and use Eq.\ (\ref{ref-dmat}).  The result is
\begin{equation} \alpha _{\gamma \eta }=\left( \varepsilon _m-1
\right)R^3 \left({ \hat {\bf e}_\gamma \bullet {\bf r} \over R} \right)
\left( {{\bf r} \bullet \hat {\bf e}_\eta \over R } \right)
.\label{app-pol} \end{equation} The exact result is \begin{equation}
\alpha _{\gamma \eta }=\left( {{{\varepsilon _m-1} \over {\varepsilon
_m+2}}} \right)R^3\delta _{\gamma \eta }.\label{exam-pol} \end{equation}
The trace of the approximate result is in error by a factor of
$3/(\varepsilon_m+2)$ that is natural.  But the approximate result is
crudely non-spherical.  This property thus provides a tougher test of
the numerical solution than does the solvation free energy because this
property depends more sensitively on the uniformity of the sampled
points.  This property is therefore helpful in discriminating different
solutions, {\it e.g.\/}, with different levels of sampling.

\subsection{More Thorough Sampling for Integration over Plaques} One way
to go beyond the method described above to obtain more accurate
computations is to evaluate the matrix elements more accurately.  This
step in achieving higher accuracy does not increase the dimension of the
set of linear equations that must be solved; solving the linear
equations is the computational limitation of the method above.  We note
further that the integrands are singular and, for typical molecular
surfaces considered, discontinuous.  These latter qualities are expected
to make these integrals tough candidates for higher-order integration
schemes.

The method described here defines plaques, as suggested above, and then,
in order to simplify the problem, assumes that one of the integrand
factors can be treated as slowly varying over each plaque.  Such an
assumption is satisfactory when the numbers of plaques are large and,
consequently, the spatial extent of each plaque is small.  Thus, we can
be sure that the numerical solution converges to an accurate result as
the number of plaques tends to infinity, just as with the method above.
However, we expect the rate of convergence of these subsequent methods
to be superior.

We begin the development by defining plaques.  The plaques will be
defined as the Voronoi polyhedra of the plaque points on the non-buried
surface of the each sphere.
Typically the
boundaries of the plaques will include the boundaries of intersection of
spheres.  The plaque points might be constructed with one of the
`sampling' methods mentioned above.  For the present development it is
not required that the method of construction of the plaque points be one
that might be used for direct estimation of the integrals.  It is
natural though to include uniformity of distribution of the plaque
points plus uniformity of size and shape of the plaques, and
compactness, as {\it desiderata\/}.  As specific examples, we have
satisfactorily used both quasi-random number series as a sampling for
the spherical surface, and the Lubotzky-Phillips-Sarnak
\cite{hammersley,keng,lubotzky,tichy,niederreiter,press} spherical
lattice to construct plaque points.  We will denote the set of plaque
points by $\{\alpha\}$.

In addition to the plaque points $\{\alpha\}$ we will require another
larger set of points that sample the molecular surface uniformly and
will be used to accomplish integrations over the plaques.  We have
typically used Monte Carlo methods with pseudo-random numbers to
generate this set of points that will be denoted by $\{i\}$.  When the
plaque points are constructed to be reasonably uniformly distributed on
the surface --- even though only a finite set of points are used --- we
would typically include those plaque points in the set $\{i\}$.

The present method can then be described by consideration of Eq.\
(\ref{inteqa}).  We assume that the $G({\bf r}'', {\bf r}')$ appearing
in the integral there is accurately a constant on each plaque.  The
equations for the off-diagonal coefficient $w_{\alpha \beta}$ that
follow from this assumption are \begin{equation} w_{\alpha \beta} =
{R(s_\beta)^2 { (\varepsilon_s - \varepsilon_m) }\over M(s_\beta)}
\sum_{\{i\in \beta \}} { ( {\bf r}_\alpha - {\bf r}_i ) \bullet {\hat
{\bf n} }_i \over |{\bf r}_\alpha - {\bf r}_i |^3 } , \qquad \alpha \ne
\beta .  \label{cor-od} \end{equation} The set of points $i$ that are
within the plaque $\beta$ is then denoted by $\{i\in \beta\}$.
$M(s_\beta)$ is the number of points, sampling points plus the plaque
point, on the sphere that supports plaque $\beta$.  Formulae for the
diagonal element $ w_{\alpha \alpha}$ that follow from the assumption
that $G({\bf r}'', {\bf r}')$ is constant over a plaque are
\begin{equation} \Omega _\alpha / 4 \pi \approx {1 \over 2\sqrt 2}\left(
{1 \over M(s_\alpha)} \right)\sum\limits_{\left\{ {i\notin \alpha}
\right\}} {\left( {1-\cos \vartheta _{i\alpha}} \right)^{-1/2}} ,
\label{fsa} \end{equation} and \begin{equation} w_{\alpha\alpha} =
\left( {\Omega_\alpha \over 4 \pi }\right) (\varepsilon_s -
\varepsilon_m ) .  \label{cor-dmat} \end{equation} The derivation of
this result is depicted in Fig.~\ref{fsa.fig} and discussed in an
appendix.  As indicated in detail by Eq.\ (\ref{ref-dmat}), this applies
to the case that the plaque point $\alpha$ is exposed.  In Eq.\
(\ref{fsa}), $\vartheta_{i\alpha}$ is the angle between the surface
normals at point $\alpha$, the plaque point, and the sampled point $i$.
That formula arranges to use sample points {\it outside\/} the plaque
$\alpha$ to calculate the solid angle subtended by that plaque at that
plaque point.  The purpose of that arrangement is to reduce the variance
of that Monte Carlo estimation.  Note that all sample points, buried or
not, on the surface of the sphere that supports plaque $\alpha$ should
be used in the estimation but whether any sample point resides within or
without plaque $\alpha$ depends on resolution of buried surface.
Eqs.~(\ref{cor-od}), (\ref{fsa}), and (\ref{cor-dmat}) are remarkably
similar to Eqs.\ (\ref{p-odmat}) and (\ref{ref-dmat}) and remarkably
simple.

\subsection{Refinement for higher resolution}

The left subscript index of $w_{\alpha \beta}$ indicates a point on the
molecular surface where the electrostatic potential is sought.  It is
straightforward to evaluate these matrix elements for points additional
to the plaque points.  It is straighforward to evalute the bare
potential at additional points also.  Evaluation of those quantities
permits a simple calculation of a refined approximate $\Phi({\bf r})$
once a preliminary or coarse solution, call it $\overline \Phi({\bf
r})$, is obtained.  We `iterate once' Eq.~(\ref{p-phidis}) with the
preliminary solution used on the right:  \begin{equation}\varepsilon_s
\Phi({\bf r}_\alpha) = \Phi^{(0)}({\bf r}_\alpha) + \sum_\beta w_{\alpha
\beta} \overline\Phi({\bf r}_\beta).  \label{corfin} \end{equation}

	In pushing to higher resolution, it is particularly helpful to
note the requirement $\sum\limits_\beta ^{} {w_{\alpha \beta }}=0$.
This follows from Gauss's law applied to Eq.~(\ref{inteqa}) when ${\bf
r}$ is outside the molecular volume.

\subsection{Coarsened equations}

	A closed set of coarsened equations can be obtained from
Eqs.~(\ref{corfin}) by the following argument:  In the calculation of
the matrix element $w_{\alpha \beta}$ it was assumed that the potential
was constant over the plaque $\beta$.  It is therefore conceptually
consistent to consider $\overline \Phi({\bf r}_\beta)$ to be the average
of the potential over the plaque rather than the value of the potential
at one central point, the plaque point ${\bf r}_\beta$.  With this
choice of $\overline \Phi({\bf r})$ as the plaque average, the plaque
average of the finer resolution Eq.~(\ref{corfin}) then gives a
coarse-grained equation that is closed in these plaque averaged
quantities:  \begin{equation}\varepsilon_s \overline\Phi({\bf r}_\alpha)
= \overline\Phi^{(0)}({\bf r}_\alpha) + \sum_\beta \overline w_{\alpha
\beta} \overline\Phi({\bf r}_\beta).  \label{corcor} \end{equation} Use
of this equation prior to the higher resolution refinement
Eq.~(\ref{corfin}) has the advantage that the coarse solution will then
already satisfy the refinement in the plaque mean.  The magnitude of the
correction due to higher resolution refinement is then reduced.

	There is an alternative to this procedure that is typically
advantageous, sometimes computationally simpler though requiring more
memory, but also conceptually more indirect.  We arrive at this
alternative procedure by the following argument:  we formulate a high
resolution set of discrete linear equations using Eqs.~(\ref{cor-od}),
(\ref{fsa}), and (\ref{cor-dmat}).  We then appeal to a coarser
tesselation of the molecular surface and require that all the points of
the fine lattice that reside on the same coarse plaque have the same
potential.  This requirement produces an overdetermined set of linear
equations that can be analyzed with the traditional application of
singular value decomposition \cite{NR} to produce the solution that
minimizes the mean square residual of fine resolution equations.  This
approach attempts to utilize the fine-scale spatial variation of the
$\Phi^{(0)}({\bf r}_\alpha)$ in constructing a coarse solution.

\subsection{Surface Charges}

An alternative implementation of Eq.\ (\ref{5}) holds an advantage for
applications of traditional electronic structure packages \cite{estruc}.
That advantage derives from the fact that the induced electrostatic
fields are represented with surface distributions of charge.
Traditional electronic structure packages can use collections of charge
monopoles simply.  If the surface charge distributions are represented
by a collection of monopoles then the utilization of traditional
electronic structure packages is logistically simplified.  In this
section we describe how the ideas above are adapted to utilization of
Eq.\ (\ref{5}) in those applications.

As with the development above we first specialize Eq.~(\ref{5}) to
extract a boundary integral equation.  This is discussed in further
details in the appendix.  We evaluate formally the gradient of that
equation and then consider how to interpret the results for
$\varepsilon({\bf r})$ a sharp step at the molecule surface.  We
interpret integrand factors that become discontinuous in that
circumstance as the average of the values inside and outside that
discontinuity.  We then take the observation point ${\bf r}$ just inside
the molecular surface and notice that we obtain the simple closed
equation \begin{equation} \hat {\bf n}\bullet {\bf D}({\bf r})=\hat {\bf
n}\bullet {\bf E}_0({\bf r})+\int\limits_S {\hat {\bf n} \bullet \nabla
G_0({\bf r},{\bf r'})\left({\varepsilon_s - \varepsilon_m \over 4\pi
\varepsilon_s } \right)\hat {\bf n}\bullet {\bf D}({\bf
r'})d^2r'}\label{Deq} \end{equation} for the Maxwell displacement normal
to the molecular surface at the point ${\bf r}$.  $E_n^0({\bf r})$ is
the component of the bare electric field --- due to $\rho$ --- normal to
the molecular surface at the point ${\bf r}$.  Eq.~(\ref{Deq}) involves
a surface integral only because it is already specialized to treat a
sharp surface.  As noted above, this equation provides the basis for the
most widely used previous boundary element approaches for these problems
\cite{rashin-review,miertus,zauhar_a,rashin_a,zauhar_b,yoon-lenhoff,%
juffer,wang}.  The discretized version of Eq.~(\ref{Deq}), comparable to
Eq.~(\ref{p-phidis}), is \begin{equation} D_n({\bf r}_i)= E_n^0({\bf
r}_i) + \sum_j w_{ij}D_n ({\bf r}_j) ,\label{p-Ddis} \end{equation}
where the off-diagonal coefficients are given by \begin{equation}
w_{ij}=-\ {{\hat {\bf n} }_i \bullet ({\bf r}_i - {\bf r}_j ) \over
|{\bf r}_i - {\bf r}_j |^3} \times { (\varepsilon_s -\varepsilon_m )
\over 4 \pi \varepsilon_s } \times \left( {4\pi R(s_j)^2 \eta({\bf
r}_j)\over M(s_j)} \right), i \ne j .\label{p-Dod} \end{equation} The
diagonal contributions can be obtained by noting that when the plaques
are small, their curvature can be neglected and that, since the plaque
point is immediately inside the plaque, edge effects may be neglected in
a first approximation.  The integration required in Eq.\ (\ref{Deq}) is
then {\it minus\/} the electric field due to a planar sheet with charge
density $\left(\varepsilon_s - \varepsilon_m\right) D_n({\bf r}) / 4\pi
\varepsilon_s$.  This electric field just inside has magnitude $2 \pi
\times$ (that surface charge density) and is anti-parallel to the
outward directed surface normal.  Thus \begin{equation} w_{jj} =
(\varepsilon_s - \varepsilon_m ) \eta({\bf r}_j)/2 \varepsilon_s .
\label{p_Ddmat} \end{equation} The refinement corresponding to Eq.\
(\ref{ref-dmat}) is here \begin{equation} w_{jj} = (\varepsilon_s -
\varepsilon_m ) (1 - M(s_j)^{-1/2})\eta({\bf r}_j)/2 \varepsilon_s .
\label{ref-Ddmat} \end{equation} After $D_n$ is obtained on the
molecular surface, we review the antecedents of Eq.\ (\ref{Deq}) to see
how to obtain the thermodynamic quantity $\Delta \mu$ of Eq.\
(\ref{dmu}).  When all the free charges are in regions where
$\varepsilon({\bf r}) = \varepsilon_m$, the induced potential should be
calculated as \begin{equation} \Phi ({\bf r})-\Phi ^{(0)}({\bf r})= -
\int\limits_S {G^{(0)}({\bf r},{\bf r}') \left({\varepsilon_s -
\varepsilon_m \over 4\pi \varepsilon_s \varepsilon_m } \right) \hat {\bf
n}\bullet {\bf D}({\bf r'}) d^2r'}, \label{induced} \end{equation} in
close similarity to Eq.\ (\ref{5}).  Again, the refinement of Eq.\
(\ref{ref-Ddmat}) produces the exact answer for the simplest problem,
the spherical ion with charge at the center mentioned above, when the
surface is sampled at one point only.

\subsection{Coarse and Finer Solutions with Surface Charges}

The arguments above that identified a higher resolution approximate and
the `self-consistent' coarse equations carry through for the present
surface charge of the adjoint prespective also.

\section{Molecular theoretical perspective}

Here we note some points of prespective gained by an appreciation of how
the dielectric models for solvation free energies connect to molecular
theory.  Some of these points have been noted before
\cite{phg,tp,lbk,wilfred,fbl,kw,hpga,hpgb}.  But they seem not be widely
appreciated.  Further, new points here clearly indicate how to improve
simply the predictions of potentials of mean force while utilizing the
methods detailed above.

The molecular theory corresponding to these dielectric models is the
second-order cumulant approximation \begin{equation} \Delta \mu \approx
\Delta \mu _0 + \left< \sum_j \varphi({\bf j}) \right> _0 -{\beta \over
2} \left< {\left( \sum_j \varphi ({\bf j}) - \left< \sum_k \varphi ({\bf
k}) \right> _0 \right)}^2 \right> _0 .  \label{perturb} \end{equation}
Here $ \varphi({\bf j}) $ is the electrostatic interaction potential
energy coupling the solute to solvent molecule {\bf j}.  The brackets
$\left< \cdots \right>_0$ indicate the thermal average in the absence of
those electrostatic couplings and $\Delta \mu_0$ is the excess chemical
potential of the solute at infinite dilution again in the absence of
electrostatic interactions.  When the solute is an infinitely dilute
second component this molecular approximation is perturbation theory
through second order in the electrostatic interactions.  When the
molecule is not literally an infinitely dilute second component the
`infinite dilution' restriction means that one molecule is distinguished
for the purposes of calculation.  This is still a natural theory but the
medium now contains a non-zero concentration of molecules mechanically
identical to the `solute.'  The medium properties non-perturbatively
reflect that fact.  From the perspective of this molecular theory, the
dielectric solvation model application neglects the zeroth and first
order terms, makes an estimate of the second-order term, and neglects
all succeeding contributions.  Refs.  \cite{lbk,wilfred,fbl} can be
consulted for alternative views.

In view of the generally satisfactory predictions of the dielectric
models in the examples above, it is reasonable to conclude that the
approximate molecular theory of Eq.~(\ref{perturb}) would provide a
valid description of those systems also but with further molecular
detail \cite{hpga,hpgb} and without the empiricism encoded in the
radii-parameters of the dielectric model.  That empiricism does in fact
limit the ability to use the model for prediction.  In the examples of
the inter-ionic potentials of mean force above, it is clear that the
dielectric model is not satisfactorily accurate.  But even then, it is a
reasonable view that the failures of the model are due principally to
failure in suitably describing the molecular volume rather than an
essential failure of the perturbative formula Eq.~(\ref{perturb}).

A direct implementation of Eq.~(\ref{perturb}) would require simulation
calculation.  Such calculations are likely only to be somewhat simpler
than the direct simulation of the potential of mean force sought.  Thus,
to extend the theoretical insight that Eq.~(\ref{perturb}) is physically
valid, other tricks must be found.  One additional idea is to attempt to
calculate the first-order and second-order terms in the molecular theory
on the basis of the dielectric models.  Direct use of that idea is
likely to be unsuccessful because those terms in the molecular theory
require analysis of the solvation of a discharged solute and the
dielectric model is unlikely to be accurate for discharged solutes.
However, a physically equivalent but operationally distinct expression
of that idea is \cite{yu-karplus,king} \begin{equation} \Delta \mu
\approx \Delta \mu _0 + \left< \sum_j \varphi({\bf j}) \right> _{1/2} .
\label{half} \end{equation} With this formula, the average can be
evaluated by considering {\it one\/} solvent molecule, a probe molecule,
in addition to the solute discharged to the extent $1/2$.  This
molecular grouping can be studied on the basis of the dielectric model
with the expectation that the dielectric model should give a physically
valid description.  The dielectric model can provide a pair potential of
mean force for the additional solvent molecule relative to the
half-discharged solute.  The mean electrostatic potential that the water
molecule expresses on the solute is then the quantity of interest.  To
obtain that quantity requires averaging over the Boltzmann factor of the
dielectric model potential of mean force for the grouping.

The justification of this approach is that the formula Eq.~(\ref{half})
would give precisely the same answer as Eq.~(\ref{perturb}) if that
latter result were precisely correct.  However, if we are looking for
one such formula to exploit with the dielectric model, Eq~(\ref{half})
is preferred because:  (a) it uses the dielectric model in a region
where it might be expected to be physically valid; and (b)
Eq~(\ref{half}) requires considering only {\it one\/} water molecule in
addition to the solute rather than two as would the last term of
Eq~(\ref{perturb}).

More formally, we can argue as follows.  The desired excess chemical
potential can be expressed as \begin{equation} \Delta \mu = \Delta \mu
_0 + \int_0^1 \left< \sum_j \varphi({\bf j}) \right> _{\xi} d\xi .
\label{f-charging} \end{equation} $\xi \in [0,1]$ is a parameter that
scales all the partial charges of the solute.  The mean value theorem
applied to the integral yields \begin{equation} \Delta \mu = \Delta \mu
_0 + \left< \sum_j \varphi({\bf j}) \right> _{\xi^\ast } ,
\label{mean_value} \end{equation} where $0<\xi^\ast<1$.  The success of
the dielectric model suggests that we use the estimate $\xi^\ast \approx
1/2$ and evaluate the average by exploiting the dielectric model itself.
This more formal argument obviously further suggests that the precise
value used for $\xi^\ast$, though near $1/2$, can be helpfully
empiricized as sufficient experience is acquired with these methods.

	In closing this section it is helpful to note some often
mentioned alternatives for the molecular theory corresponding to this
dielectric model.  One alternative might be based upon the suggestion
that a spatially local dielectric constant should be used.  If a first
solvation shell is structurally fixed and if a local dielectric constant
is given to describe the polarization of that shell, then this is
similar in spirit to the present model.  However, more generally it must
be admitted that a local dielectric constant is a ambiguous concept
unless local means `macroscopically local.'  The macroscopic solution
dielectric constant is unambiguous and the net polarization of the
interior of a rigid molecular structure also is a simple physical
concept.  This level of modeling is therefore reasonable but we have
refrained from extending the present model to elaborate those ideas in
more difficult settings.

	A second alternative is based upon the observation that on a
molecular scale the dielectric response should be non-local.  This
non-local response would be characterized by a wave-vector dependent
dielectric constant, $ \hat \varepsilon (k) $.  This is a much less
ambiguous approach than the `local dielectric constant' idea mentioned
above.  However, it does not go as far as the second-order perturbation
theory advocated here.  The `non-local' response approach would not
typically include the linear perturbative term.  The non-local response
approach would typically compromise the description of short-ranged
`packing' interactions as they are expressed in the perturbative terms.
Moreover, the $ \hat \varepsilon (k) $ is not necessarily simple
\cite{bagchi,kirzhnits}.

\section{Conclusions}

This dielectric model \cite{rashin-review,honig-review} gives
interesting and helpful results for the variation in solvation free
energy with solute geometry.  However, it typically significantly
over-stabilizes classic attractive ion-pairing configurations.

In several cases, {\it e.g.\/} Fig.~\ref{methchl.fig} and
Fig.~\ref{fig.nmaa}, the predictions of this dielectric model are quite
similar to those of more ambitious theories.  This comparison clearly
teaches us about the performance of those more complicated theories for
those cases:  the phenomena described are unsuspectedly simple and those
more complicated theories are at least qualitatively valid for those
simple circumstances.

It is numerically feasible to solve this dielectric model to thermal
accuracies for the excess chemical potential
\cite{simonson,tucker,bharadwaj}.  The integral equations
Eqs.~(\ref{inteq}) and (\ref{refinteq}) that are the basis for the
numerical boundary integral methods studied here have not been given
previously in this context.  The algorithmic discussion of
Eq.~(\ref{inteq}) has laid the basis for a multigrid method of solution.
Such methods are likely to be necessary for similarly accurate numerical
solution of these models for much larger solutes.  Eq.~(\ref{refinteq})
will be particularly helpful for macroscopic surfaces such as solution
interfaces and membranes.  Eqs.~(\ref{b3}), (\ref{b8}), and (\ref{b9})
show how these methods can be transferred directly to periodic boundary
conditions.  That will be important for two reasons.  First, the best
controlled quantitative results on hydration and electrostatic
interactions utilize periodic boundary conditions on electrostatic
interactions.  Second, simulation of the hydration of macromolecules
often involves periodic boundary conditions, only a few layers of
hydrating water, and almost no direct investigation of system size
effects; treatment of hydration on the basis of the dielectric model
with and without periodic boundary conditions would be a natural first
step towards getting quantitative thermodynamic limiting results for
such solutions.

Finally, we draw attention to the connection discussed here of this
dielectric model to molecular theory.  As a proposal for going beyond
current applications of dielectric models, the probe solvent molecule
Eq.~(\ref{half}) is new.  By reintroducing molecular solvation
structure, it is likely to correct the largest part of the errors of the
dielectric model, {\it i.e.\/}, the quantitative inaccuracy for either
molecular surface used of the contact minimum in Fig.~\ref{nacl.fig} and
of the barrier to escape from that contact minimum.  As a guide in
analyzing molecular statistical thermodynamic results, this relation has
already been useful.  At a more practical level, this approach should
reduce the sensitivity of the model results on the radii-parameters.  It
should be cautioned however, that the dielectric model is clearly
approximate and the quadratic theory also is approximate
\cite{hpga,hpgb}.  Typically, these results can be expected to be
superceded for applications where molecular accuracy is important.

\appendix{{\bf Acknowledgements}}

This work has benefitted from a helpful collaboration with
scientists at 3M Corporation lead by David Misemer, and from discussions
with a number of colleagues Richard Friesner, Barry Honig, Ron Levy,
Alex Rashin, and Andrew Pohorille.  This work has been supported by the
Tank Waste Remediation System (TWRS) Technology Application program,
under the sponsorship of the U.  S.  Department of Energy EM-36, Hanford
Program Office, the Air Force Civil Engineering Support Agency, and the
US-DOE under LANL Laboratory Directred Research and Development funds.

\bigskip
\appendix{{\bf Appendix A: Integral formulations}}

We wish to write the equations to be solved as integral equations
because those integral equations serve as the basis of the numerical
methods here and because this is the most natural way to make contact
with molecular theory.  We consider the equation for the Green function,
$G({\bf r}, {\bf r}')$:  \begin{equation} \nabla \bullet {\varepsilon
({\bf r})\nabla G({\bf r},{\bf r}')} =-4\pi \delta ({\bf r}-{\bf r}') .
\label{a1} \end{equation} $G({\bf r}, {\bf r}')$ is the potential at
${\bf r}$ due to a unit charge at ${\bf r}'$.  We can rearrange this
into a perturbation theory standardly.  Consider the reference problem
\begin{equation} \varepsilon_0 \nabla \bullet \nabla G^{(0)}({\bf
r},{\bf r}') =-4\pi \delta ({\bf r}-{\bf r}') .
\label{a2}\end{equation} $G^{(0)}({\bf r}, {\bf r}')$ is the potential
at ${\bf r}$ due to a unit charge at ${\bf r}'$ for a uniform dielectric
with dielectric constant $\varepsilon_0$.  Then the perturbation theory
is \begin{equation} G({\bf r},{\bf r}') =G^{(0)}({\bf r},{\bf
r}')+\int\limits_V {G^{(0)}({\bf r},{\bf r}'') \nabla ''\bullet \left[
{\left( {{{\varepsilon ({\bf r}'')-\varepsilon _0} \over {4\pi }}}
\right)\nabla '' G({\bf r}'',{\bf r}')} \right]d^3r''} .
\label{a3}\end{equation} When $\varepsilon_0 = 1$ we interpret this
equation by noting that \begin{equation} \nabla ''\bullet \left[ {\left(
{{{\varepsilon ({\bf r}'')-\varepsilon _0} \over {4\pi }}} \right)\nabla
''G({\bf r}'',{\bf r}')} \right] \end{equation} is then the induced
charge at ${\bf r}''$ due to a unit charge at ${\bf r}'$.  The total
potential $G({\bf r}, {\bf r}')$ at ${\bf r}$ is the sum of the bare
potential $G^{(0)}({\bf r}, {\bf r}')$ of the unit source charge plus
the bare potential of all the induced charges.

An integration by parts casts this into the form:  \begin{equation}
G({\bf r},{\bf r}')=G^{(0)}({\bf r},{\bf r}') -\int\limits_V {\left[
{\nabla ''G^{(0)}({\bf r},{\bf r}'')} \right]\bullet \left(
{{{\varepsilon ({\bf r}'')-\varepsilon _0} \over {4\pi }}}\right) \left[
{\nabla ''G({\bf r}'',{\bf r}')} \right]d^3r''} .  \label{a4}
\end{equation}

The surface contributions vanish because we are here considering a
charge distribution of finite extent and the potential is required to
vanish at infinity.  This form makes evident the well-known symmetry of
$G({\bf r}, {\bf r}')$.  Now when $\varepsilon_0=1$ the interpretation
is that
\begin{equation} \left({{{\varepsilon ({\bf r}'')-\varepsilon
_0}\over{4\pi }}}\right) \left[{-\nabla ''G({\bf r}'',{\bf r}')} \right]
\end{equation}
is the induced polarization --- $4\pi {\bf P}=(\varepsilon
-1){\bf E}$.  Since $\nabla ''G^{(0)}({\bf r},{\bf r}'')\bullet {\bf u}$ is
the potential at ${\bf r}$ due to a unit dipole ${\bf u}$ at ${\bf
r}''$, the total potential $G({\bf r}, {\bf r}')$ at ${\bf r}$ is the
sum of the bare potential $G^{(0)}({\bf r}, {\bf r})$ of the unit source
charge plus the bare potential of all the induced polarization.

The form \begin{equation} G({\bf r},{\bf r}') =G^{(0)}({\bf r},{\bf
r}')+\int\limits_V {\left\{ {\nabla ''\bullet \left[ {\left(
{{{\varepsilon ({\bf r}'')-\varepsilon _0} \over {4\pi }}} \right)\nabla
''G^{(0)}({\bf r},{\bf r}'')} \right]} \right\}G({\bf r}'',{\bf
r}')d^3r''} \label{a5} \end{equation} is obtained by another integration
by parts.  By elaborating the gradient operations we see that
\begin{equation} \nabla ''\bullet \left[ {\left( {{{\varepsilon ({\bf
r}'')-\varepsilon _0} \over {4\pi }}} \right)\nabla ''G^{(0)}({\bf
r},{\bf r}'')} \right] =\left[ {\nabla ''\left( {{{\varepsilon ({\bf
r}'')} \over {4\pi }}} \right)} \right]\bullet \nabla ''G^{(0)}({\bf
r},{\bf r}'')-\left( {{{\varepsilon ({\bf r}'')-\varepsilon _0} \over
{\varepsilon _0}}} \right)\delta ({\bf r}-{\bf r}'').  \label{a6}
\end{equation} Our basic equation then becomes \begin{equation}
\varepsilon ({\bf r})G({\bf r},{\bf r}') = \varepsilon _0G^{(0)}({\bf
r},{\bf r}') +\int\limits_V {\left[ {\nabla ''\varepsilon _0G^{(0)}({\bf
r},{\bf r}'')} \right]\bullet \left[ {{{\nabla ''\varepsilon ({\bf
r}'')} \over {4\pi \varepsilon ({\bf r}'')}}} \right]\left[ {\varepsilon
({\bf r}'')G({\bf r}'',{\bf r}')} \right]d^3r''} .  \label{a7}
\end{equation} $\varepsilon _0G^{(0)}({\bf r},{\bf r}')$ is the bare
Green function for a uniform dielectric with dielectric constant one.
Thus we write \begin{equation} \varepsilon({\bf r})G({\bf r},{\bf r}')=
G^{(0)}({\bf r},{\bf r}') +\int\limits_V { {\nabla ''G^{(0)}({\bf
r},{\bf r}'')} \bullet \left[ {{{\nabla ''\varepsilon ({\bf r}'')} \over
{4\pi \varepsilon ({\bf r}'')}}} \right] {\varepsilon ({\bf r}'')G({\bf
r}'',{\bf r}')} d^3r''} , \label{a8} \end{equation} with the
understanding that here $G^{(0)}({\bf r},{\bf r}')$ is the vacuum Green
function.  This integral form of Poisson's equation directly gives the
correct answer for a uniform dielectric.  It also admits a simple
interpretation.  $\left[ {\nabla ''\varepsilon ({\bf r}'') / 4\pi }
\right]G({\bf r}'',{\bf r}') $ can be viewed as a surface density of
dipoles.  The remaining integrand factor $ \nabla'' G^{(0)}({\bf r},{\bf
r}'') $ is then the potential due to those surface dipoles.  This
formula has the following consistency:  the indirect term on the
right-side, that associated with the surface dipole distribution, is
discontinuous across the surface.  The value of such a discontinuity is
$4 \pi \times $ (the local surface density of dipoles).  Here that is $4
\pi [(\varepsilon -1)/ (4 \pi)] G({\bf r}, {\bf r}')$ for {\bf r}
located on the surface.  The left-side of this equation also is
discontinuous and the amount of that discontinuity is $(\varepsilon -1)
G({\bf r}, {\bf r}')$.

If we manipulate the Eq.~(\ref{a3}) this way we obtain
\begin{eqnarray} \nabla ''\bullet \left[ {\left( {{{\varepsilon ({\bf
r}'')-\varepsilon _0}}} \right)\nabla ''G({\bf r}'',{\bf
r}')} \right] & = &\left[ {\nabla
''\varepsilon ({\bf r}'')} \right]\bullet \left[ {\nabla ''G({\bf
r}'',{\bf r}')} \right] \nonumber \\ & + & \left( {{{\varepsilon ({\bf
r}'')-\varepsilon _0} \over {\varepsilon ({\bf r}'')}}}
\right)\left\{ {-4\pi \delta ({\bf r}''-{\bf r}')-\left[ {\nabla
''\varepsilon ({\bf r}'')} \right]\bullet \left[ {\nabla ''G({\bf
r}'',{\bf r}')} \right]} \right\}.  \label{a9}\end{eqnarray}
Thus
\begin{equation} G({\bf r},{\bf r}')\varepsilon ({\bf r}')=G^{(0)}({\bf
r},{\bf r}') +\int\limits_V G^{(0)}({\bf r},{\bf r}'') \left[ {\nabla
''\varepsilon ({\bf r}'')\over 4\pi\varepsilon ({\bf r}'')} \right]
\bullet {\nabla ''G({\bf r}'',{\bf r}')\varepsilon ({\bf r}') d^3r''} .
\label{a10} \end{equation}
Here again we understand that $G^{(0)}({\bf r},{\bf r}')$ is the vacuum
Green function.  This equation can be viewed as the basis of the
previous boundary element numerical approaches
\cite{rashin-review,miertus,zauhar_a,rashin_a,zauhar_b,yoon-lenhoff,%
juffer,wang}.  If the sources are located at positions ${\bf r}'$ where
the dielectric function $\varepsilon({\bf r}')$ is {\it one\/} then this
equation states that the full potential can be composed as the bare
potential of those sources plus the bare potential due to charges
distributed throughout regions of inhomogeneity of $\varepsilon({\bf
r})$.

To show the explicit correspondence with previous workers, we assume
that all charges are located in regions for which $\varepsilon({\bf
r}) = 1$, right-sum over those charges, operate on Eq.~(12) with
$-\hat {\bf n} \bullet \nabla$ for ${\bf r}$ just inside the molecular
surface to obtain
\begin{equation} \hat {\bf n}\bullet {\bf D}({\bf r})=\hat {\bf
n}\bullet {\bf E}_0({\bf r})+\int\limits_S {\hat {\bf n} \bullet \nabla
G^{(0)}({\bf r},{\bf r'})\left( {{{\varepsilon -1} \over {4\pi
\varepsilon }}} \right)\hat {\bf n}\bullet {\bf D}({\bf r'})d^2r'}
.\label{a11} \end{equation}
The right-most quantity in the integrand --- the surface density of
charge --- is worked out as:

\begin{eqnarray} q({\bf r'}) &= &\left( {\varepsilon -1 \over 4\pi }
\right) \left( {2 \over \varepsilon +1} \right)
({1 \over 2})
(\hat{\bf n}\bullet {\bf D}({\bf r'}) / \varepsilon +\hat {\bf n}\bullet
{\bf D}({\bf r'})))
 \nonumber \\ & = &\left( {\varepsilon -1 \over 4\pi
\varepsilon}\right) \hat {\bf n} \bullet {\bf D} ({\bf r'}) \label{a12}
\end{eqnarray}

These integral formulations can be generalized to cases in which the
zeroth order contribution is a solution for a similar problem with
nonconstant $\varepsilon({\bf r})$.  Such generalizations will be
helpful in treating molecules near surfaces.  The required developments
here follow the above work closely.  We reconsider the reference problem
and select \begin{equation} \nabla \bullet \varepsilon_0({\bf r}) \nabla
G^{(0)}({\bf r},{\bf r}') =-4\pi \delta ({\bf r}-{\bf r}') .
\label{a14} \end{equation} $G^{(0)}({\bf r}, {\bf r}')$ is the potential
at ${\bf r}$ due to a unit charge at ${\bf r}'$ for the reference
problem with dielectric constant $\varepsilon_0({\bf r})$.  Then the
perturbation theory is \begin{equation} G({\bf r},{\bf r}')
=G^{(0)}({\bf r},{\bf r}') +\int\limits_V G^{(0)}({\bf r},{\bf r}'')
\nabla ''\bullet \left[ \left( {\Delta \varepsilon ({\bf r}'') \over
4\pi } \right)\nabla '' G({\bf r}'',{\bf r}') \right]d^3r'' \label{a15}
\end{equation} where we use the notation $\Delta \varepsilon ({\bf r})=
\varepsilon ({\bf r})-\varepsilon _0({\bf r})$.  We then proceed in the
obvious way:  \begin{equation} G({\bf r},{\bf r}') =G^{(0)}({\bf r},{\bf
r}') +\int\limits_V G^{(0)}({\bf r},{\bf r}'') \nabla ''\bullet \left[
\left( {\Delta \varepsilon ({\bf r}'') \over 4\pi \varepsilon ({\bf
r}'')} \right) \varepsilon ({\bf r}'')\nabla '' G({\bf r}'',{\bf r}')
\right]d^3r'' .  \label{a16} \end{equation} Working out the $\nabla$'s
gives \begin{equation} G({\bf r},{\bf r}') =G^{(0)}({\bf r},{\bf r}')
\varepsilon_0({\bf r}')/\varepsilon({\bf r}') +\int\limits_V
G^{(0)}({\bf r},{\bf r}'') \nabla ''\left( {\Delta \varepsilon ({\bf
r}'') \over 4\pi \varepsilon ({\bf r}'')} \right)\bullet \varepsilon
({\bf r}'')\nabla '' G({\bf r}'',{\bf r}') d^3r'' .  \label{a17}
\end{equation} But \begin{equation} \varepsilon({\bf r}'') \nabla''
\left( {\Delta \varepsilon ({\bf r}'') \over \varepsilon({\bf
r}'')}\right) = \varepsilon_0({\bf r}'') \nabla'' \ln \left(
{\varepsilon ({\bf r}'') \over \varepsilon_0 ({\bf r}'')} \right) .
\label{a18} \end{equation} Then finally \begin{eqnarray} G({\bf r},{\bf
r}') \varepsilon({\bf r}') &=&G^{(0)}({\bf r},{\bf
r}')\varepsilon_0({\bf r}') \nonumber \\ &+&\int\limits_V G^{(0)}({\bf
r},{\bf r}'') \varepsilon_0({\bf r}'') \nabla'' \ln \left( {\varepsilon
({\bf r}'') \over \varepsilon_0 ({\bf r}'')} \right) \bullet \nabla ''
G({\bf r}'',{\bf r}') \varepsilon({\bf r}')d^3r'' .  \label{a19}
\end{eqnarray} This is the desired generalization of Eq.~(\ref{a10}).

	For the generalization of Eq.~(\ref{a8}), we proceed similarly
but start from the adjoint form:  \begin{equation} G({\bf r},{\bf r}')
=G^{(0)}({\bf r},{\bf r}') +\int\limits_V {\left\{ {\nabla ''\bullet
\left[ {\left( {{{\Delta \varepsilon ({\bf r}'')} \over {4\pi }}}
\right)\nabla ''G^{(0)}({\bf r},{\bf r}'')} \right]} \right\}G({\bf
r}'',{\bf r}')d^3r''} \label{a20} \end{equation} We then find
\begin{eqnarray} \varepsilon ({\bf r}) G({\bf r},{\bf r}') &=&
\varepsilon_0({\bf r}) G^{(0)}({\bf r},{\bf r}') \nonumber \\
&+&\int\limits_V \varepsilon_0({\bf r}) \nabla '' G^{(0)}({\bf r},{\bf
r}'') \bullet ({1 \over 4 \pi }) \bigl[ \nabla'' \ln \left( {\varepsilon
({\bf r}'') \over \varepsilon_0 ({\bf r}'')} \right) \bigr]
\varepsilon({\bf r}'') G({\bf r}'',{\bf r}') d^3r'' .  \label{a21}
\end{eqnarray}

\bigskip \appendix{{\bf Appendix B:  Boundary integrals, periodic
boundary conditions, and dielectric models}}

The methods above can be used also to solve these dielectric
models in periodic boundary conditions.  This may be important because
almost all simulations of macroscopic matter utilize periodic boundary
conditions; and many of those use the solution of Poisson's equation in
periodic boundary conditions, the Ewald potential.

We start by noting that the Ewald potential could be constructed
with a boundary integral method.  This can be demonstrated in a classic
fashion from Green's second formula \cite{green2} \begin{equation}
\int\limits_V {d^3r\left\{ {f( {\bf r})\nabla ^2g( {\bf r})-g( {\bf
r})\nabla ^2f( {\bf r})} \right\}} =\int\limits_{\partial V} {d^2r\,\hat
n\bullet \left\{ {f( {\bf r})\nabla g( {\bf r})-g( {\bf r})\nabla f(
{\bf r})} \right\}} \label{b1} \end{equation} in the present notation.
The integral on the right is over the surface of a unit cell of a
Bravais lattice, {\it e.g.\/}, the surface of a cube.  $\hat {\bf n}$ is
the outward pointing normal.  We exploit this in the traditional fashion
with the traditional choice $f( {\bf r}) = 1/\vert {\bf r} - {\bf r'}
\vert$ and $g( {\bf r}) = \varphi_{Ewald}( {\bf r})$.  Then $\nabla ^2f(
{\bf r}) = -4\pi \delta ({\bf r} - {\bf r'}) $ and $\nabla ^2g( {\bf r})
= -4\pi \left(\delta ({\bf r} - {\bf r'}) - V^{-1}\right)$.  This gives
the equation \begin{equation} \varphi _{Ewald}({\bf r})= {1 \over
r}-\int\limits_V {{{d^3r'} \over {V\left| {{\bf r}-{\bf r'}} \right|}}}+
\int\limits_{\partial V} {d^2r'\,\hat n\bullet \left\{ {{{\nabla
'\varphi _{Ewald}({\bf r'})} \over {\left| {{\bf r}-{\bf r'}} \right|}}
-\varphi _{Ewald}({\bf r'})\nabla '{1 \over {\left| {{\bf r'}-{\bf r}}
\right|}}} \right\}}.  \label{b2} \end{equation} This notation assumes,
and our argument depends on this construction, that the unit cell is
centered on the source.  This will not limit the generality of the
result because we will use that device to construct the general periodic
solution of interest.  With this choice of origin, the Ewald electric
field normal to the surface must be zero.  Then \begin{equation}\varphi
_{Ewald}({\bf r})={1 \over r}-\int\limits_V {{{d^3r'} \over {V\left|
{{\bf r}-{\bf r'}} \right|}}} -\int\limits_{\partial V} {d^2r'\,\hat
n\bullet \varphi _{Ewald}({\bf r'})\nabla '{1 \over {\left| {{\bf
r'}-{\bf r}} \right|}}}\label{b3} \end{equation} Thus, the Ewald
potential can be represented as the Coulomb potential due to a charge
distribution within the cell plus the Coulomb potential due to a dipolar
surface distribution.  This equation might be used to determine the
Ewald potential by discretizing the surface, taking the position ${\bf
r}$ on the surface, and using Eq.~(\ref{b3}) to determine
$\varphi_{Ewald}({\bf r})$ on the surface.  The periodic solution could
then be constructed by periodic replication.  However, this equation is
properly deficient in one respect:  it is ambiguous with respect to the
addition of a spatial constant.  Thus if $\overline \varphi({\bf r})$ is
a solution, so is $\overline \varphi({\bf r})+C$.  It is traditional
\cite{brush} and natural \cite{nijboer} to chose this constant so that
\begin{equation} \int\limits_V {\varphi _{Ewald}({\bf r})d^3r}=0.
\label{b4} \end{equation}

With this backdrop, we reconsider the dielectric problem.  A relevant
Green's formula is now \begin{eqnarray} \int\limits_V {d^3r\left\{
{f({\bf r})\nabla \bullet \left[ {\varepsilon ({\bf r})\nabla g({\bf
r})} \right] -g({\bf r})\nabla \bullet \left[ {\varepsilon ({\bf
r})\nabla f({\bf r})} \right]} \right\}} \nonumber\\
=\int\limits_{\partial V} {d^2r\,\hat n\bullet \left\{ {f({\bf
r})\varepsilon ({\bf r})\nabla g({\bf r}) -g({\bf r})\varepsilon ({\bf
r})\nabla f({\bf r})} \right\}} \label{b5} \end{eqnarray} We choose
\begin{eqnarray} f({\bf r})=\varphi _{Ewald}({\bf r}-{\bf r'}),& &\nabla
^2f({\bf r})=-4\pi \left( {\delta ({\bf r}-{\bf r'})-V^{-1}} \right),
\nonumber \\ g({\bf r})=\varphi ({\bf r},{\bf r''}),& &\nabla \bullet
\varepsilon ({\bf r})\nabla g({\bf r})=-4\pi \left( {\delta ({\bf
r}-{\bf r''})-V^{-1}} \right).  \label{b6} \end{eqnarray} The final
point to make about this application of Green's formulae is that in this
case the surface contributions will vanish:  \begin{equation}
\int\limits_{\partial V} d^2r\,\hat n\bullet \left\{ {\varphi
_{Ewald}({\bf r}-{\bf r'})\varepsilon ({\bf r})\nabla \varphi ({\bf
r},{\bf r''}) -\varphi ({\bf r},{\bf r''})\varepsilon ({\bf r})\nabla
\varphi _{Ewald}({\bf r}-{\bf r'})} \right\}=0.  \label{b7}
\end{equation} The reason for this is that all functions in the
integrand are triply-periodic.  Thus, the gradient at a surface point
can always be matched against another with equal weight and opposite
projection on the surface normal.  The relations $\nabla \bullet
\varepsilon ({\bf r})\nabla f({\bf r})= \left( {\nabla \varepsilon ({\bf
r})} \right)\bullet \left( {\nabla f({\bf r})} \right) +\varepsilon
({\bf r})\nabla ^2f({\bf r})$ and $\varphi _{Ewald}({\bf r}-{\bf r'})
=\varphi _{Ewald}({\bf r'}-{\bf r})$ then directly lead to

\begin{eqnarray} \varepsilon ({\bf r})\varphi ({\bf r},{\bf r'})&
=&\varphi _{Ewald}({\bf r}-{\bf r'})+ \left( {{1 \over {4\pi }}}
\right)\int\limits_V d^3r''\nabla '' \varphi _{Ewald}({\bf r}-{\bf
r''})\bullet \nabla ''\varepsilon ({\bf r''})\varphi ({\bf r''},{\bf
r'}) \nonumber \\ & + & V^{-1}\int\limits_V d^3r \left[ \varepsilon
({\bf r})\varphi ({\bf r},{\bf r'})-\varphi _{Ewald}({\bf r}-{\bf r'})
\right] \label{b8}.  \end{eqnarray}

This should be compared to Eq.~(\ref{a8}).  Now if we use the
conventional Ewald potential, then we have Eq.~(\ref{b4}).  Further, if
we use Eq.~(\ref{b8}) {\it without\/} the penultimate contribution, then
$\varepsilon ({\bf r'})\varphi ({\bf r'})$ is a linear combination of
just those conventional potentials.  In that case, the neglected last
term will be zero.  Our conclusion is that the form of the equation need
not change in going from the unbounded to the Ewald case provided we
substitute the conventional Ewald potential for Coulomb's law.

	A similar calculation starting from Eq.~(\ref{b1}) yields
\begin{eqnarray} \varphi ({\bf r},{\bf r'}) & = & \varphi _{Ewald}
({\bf r}-{\bf r'}) / \varepsilon ({\bf r'})+\int\limits_V d^3r'' \varphi
_{Ewald}({\bf r}-{\bf r''})\left( {\nabla ''\varepsilon ({\bf r''})
\over 4\pi \varepsilon ({\bf r''})} \right)\bullet \nabla ''\varphi
({\bf r''},{\bf r'}) \nonumber \\ & + & V^{-1}\int\limits_V d^3r\left[
\varphi ({\bf r},{\bf r''})-\varphi _{Ewald}({\bf r}-{\bf r''}) /
\varepsilon ({\bf r}) \right].
\label{b9}\end{eqnarray}

Again, if we use the conventional Ewald potential, the last term
on the right vanishes.

\bigskip
\appendix{{\bf Appendix C: Solid angle from sampled surface}}

Here we consider the formula Eq.~(\ref{fsa}).  Our goal is to show how
to use points sampled uniformly on the surface of a sphere to estimate
the solid angle subtended by a plaque at the plaque point just outside
the sphere.

Fig.  \ref{fsa.fig} shows the geometry considered here.  We take the
z-axis through the plaque point with origin at the center of the sphere.
The complication is that the solid angle required is referenced to the
origin at the plaque point.  The solid angle subtended just {\it
inside\/} the plaque point is given by \begin{equation} \Omega
_{inside}=2\pi +\int {H(\omega )d\omega } \label{c1} \end{equation}
where $d\omega$ is denotes the integration over the surface of a unit
sphere centered at the plaque point and $H(\omega)$ is the indicator
function for the angles $\omega$ for which the unit vector $\hat\omega$
points within the plaque.  This integration can be implemented as
\begin{equation} \Omega _{inside}=2\pi +\int_0^{2\pi } {d\varphi
}\int_{-1}^0 {d\left( {\cos \vartheta } \right)} H(\varphi ,\vartheta )
.  \label{c2} \end{equation} In order to use sampling points uniformly
distributed on the surface of the sphere we need to reference the
integration to the center of the sphere rather than the plaque point.
In that change of angular coordinates the azimuthal angle $\varphi$ is
unchanged but the polar angle $\vartheta$ transforms as \begin{equation}
d\left( {\cos \vartheta } \right)= d(\cos \vartheta _s)/ \left[ {2^{3/
2}\sqrt {1-\cos \vartheta _s}} \right].  \label{c3} \end{equation} The
polar angle from the center of the sphere is $\vartheta_s$.  We then
write \begin{eqnarray} \Omega _{inside}&=&2\pi +\int_0^{2\pi } {d\varphi
}\int_{-1}^0 {H(\varphi ,\vartheta )d\left( {\cos \vartheta } \right)}
\nonumber \\ &=&2\pi + 4\pi{{\int_0^{2\pi } {d\varphi }\int_{-1}^0
{H(\varphi ,\vartheta )d\left( {\cos \vartheta } \right)}} \over
{\int_0^{2\pi } {d\varphi }\int_{-1}^1 {d\left( {\cos \vartheta _s}
\right)}}}.  \label{c4} \end{eqnarray} This is in a form of an average:
\begin{equation} \Omega _{inside}=2\pi +{4 \pi \over {2^{3/
2}}}\left\langle {\left( {1-\cos \vartheta _s} \right)^{-1/ 2}}
\right\rangle _0 \label{c5} \end{equation} where we have used the
transformation Eq.~(\ref{c3}).  The subscript $0$ indicates that the
sampling points should be uniformly distributed on the surface of the
sphere.

	It can be checked that this formula produces the correct answer when the
plaque covers the entire sphere.  When the plaque is a circular cap
centered at the plaque point this formula produces
\begin{equation}
\Omega _{inside}=2\pi
\left( {1+\sqrt {\left( {1-\cos \vartheta ^*} \right)/ 2}} \right)
\label{c6}
\end{equation}
where $\vartheta^*$ is the polar angle at the perimeter of the circular
cap.  The use of the engineering estimate
\begin{eqnarray}
\left( {1-\cos \vartheta ^*} \right)/ 2
&=& 2 \pi \left( {1-\cos \vartheta ^*} \right)/ 4 \pi
\nonumber \\
&\approx& ({4 \pi \over N}) \times \left( {1\over 4 \pi}\right)
\label{c7}
\end{eqnarray}
underlies the approximation Eq.~(\ref{ref-dmat}).

The integrand of Eq.~(\ref{c5}) is integrable but singular.  This
difficulty can be relieved by considering the complementary indicator
function $1-H(\varphi,\vartheta)$ and the complementary solid angle:
\begin{eqnarray} 4\pi -\Omega _{inside}&=&\int_0^{2\pi } {d\varphi
}\int_{-1}^0 {d(\cos \vartheta )\left( {1-H(\varphi ,\vartheta )}
\right)} \nonumber \\ &=&{{4\pi } \over {2^{3/ 2}}} \left\langle {\left(
{1-H(\varphi ,\vartheta )} \right) \left( {1-\cos \vartheta _s}
\right)^{-1/ 2}} \right\rangle _0.  \label{c8} \end{eqnarray} But this
complement is just the outside solid angle desired, $\Omega_{outside} =
4\pi - \Omega_{inside}$.  Implementing this formula as a sampling
estimate that uses the points uniformly distributed on the spherical
area outside the plaque gives Eq.~(\ref{fsa}).

\begin{figure} \caption{ Theoretical results for the Na$^+ \cdots
$Cl$^-$ potential of mean force in water under physiological
conditions.  The two molecular dynamics results differ principally in
the fact that the upper molecular dynamics curve was obtained at
non-zero concentration of salt; see Ref.~\protect\cite{hummer92}.  The
curve labeled `molecular dynamics 91' is the result of
Ref.~\protect\cite{guardia91a}.  A simple translation nearly suffices
to superpose these two molecular dynamics results.  The XRISM curve is
redrawn from Ref.~\protect\cite{pr}.  `Dielectric 89' is redrawn from
Ref.~\protect\cite{rashin89} and `dielectric 94' is redrawn from
Refs.~\protect \cite{phg,tp}.  The two dielectric calculations differ
in that the earlier one used the Connolly molecular surface and the
later one used the van der Waals surface.  Those results are
accurately the same in the region of the contact minimum but the
different surfaces imply appreciably different results in the region of
the free energy barrier.}  \label{nacl.fig} \end{figure}

\begin{figure} \caption{Theoretical results for the Cl$^- \cdots
$Cl$^-$ potential of mean force in water under physiological
conditions.  The two molecular dynamics results differ principally in
the fact that the lower molecular dynamics curve was obtained at
non-zero concentration of salt (Ref.~\protect\cite{hummer92}).  The
curve labeled `molecular dynamics 91' is the result of
Ref.~\protect\cite{guardia91b}.  A simple translation nearly suffices
to superpose these two molecular dynamics results.  The XRISM curve is
redrawn from Ref.~\protect\cite{pr}.  `Dielectric 89' is redrawn from
Ref.~\protect\cite{rashin89} and `dielectric 95' is new here.  The two
dielectric calculations differ in the molecular surface assumed as
discussed in the text and the caption of Fig.~\protect\ref{nacl.fig}.}
\label{clcl.fig} \end{figure}

\begin{figure} \caption{Theoretical potentials of mean force for
approach of a Cl$^-$ ion to a planar t-butyl$^+$ ion.  See
Refs.~\protect\cite{ford,wljtb}. } \label{tbut.fig} \end{figure}

\begin{figure} \caption{Theoretical potentials of mean force for
approach of Na$^+$ along the P-O bond of an exposed oxygen atom in
methylphosphate$^-$ --- the asymmetric approach.  See
Refs.~\protect\cite{nadmpa,nadmpb,nadmpc}.} \label{nadmpasm.fig}
\end{figure}

\begin{figure} \caption{Theoretical potentials of mean force for
approach of Na$^+$ along the bisector of the O-P-O bond angle of exposed
oxygen atoms in methylphosphate$^-$ --- the symmetric approach.  See
Refs.~\protect\cite{nadmpa,nadmpb,nadmpc}}. \label{nadmpsym.fig}
\end{figure}

\begin{figure} \caption{Theoretical potentials of mean force for
exchange of Cl$^-$ in methylchloride by a symmetric S$_N$2 process
along a linear reaction path.  See
Refs.~\protect\cite{tp,ford,sn2a,sn2b,sn2c}.  The reaction coordinate
in the difference between the two carbon-chloride distances.}
\label{methchl.fig} \end{figure}

\begin{figure} \caption{Theoretical potentials of mean force for
nucleophilic addition of hydroxide to formaldehyde.  See
Refs.~\protect\cite{tp,forma,formb}.  The reaction coordinate plotted
is the carbon-oxygen distance.}  \label{formaldehyde.fig} \end{figure}

\begin{figure} \caption{Solvation contribution to the potential of mean
forces for model interconversion of {\it cis\/} [left, $\omega =
0^\circ$] to {\it trans\/} [right, $\omega = 180^\circ$].  See text and
Refs.~\protect\cite{nmaaa,nmaab,nmaac}.}  \label{fig.nmaa} \end{figure}

\begin{figure} \caption{Geometry for calculation of the solid angle
subtended at a plaque point.}  \label{fsa.fig} \end{figure}


\begin{references}

\bibitem{rashin-review} Rashin, A.  A.  {\em J.  Phys.  Chem. } {\bf
1990}, 94, 1725.

\bibitem{honig-review} Honig, B.; Sharp, K.;  Yang, A.-S. {\em J.
Phys.  Chem.} {\bf 1993}, 97, 1101.

\bibitem{miertus} Miertus, S .; Scrocco, E.;  Tomasi, J.  {\em Chem.
Phys.}  {\bf 1981}, 55, 117.

\bibitem{warwicker} Warwicker J.;  Watson, H.  C.  {\em J.  Mol.
Biol.}  {\bf 1982}, 157, 671.

\bibitem{gilson_a} Gilson, M.  K.; Rashin, A.; Fine, R.;  Honig, B.
{\em J.  Mol.  Biol.}  {\bf 1985}, 183, 503.

\bibitem{zauhar_a} Zauhar, R. J.;  Morgan, R. S. {\em J.  Mol.
Biol.} {\bf 1985}, 186, 815.

\bibitem{klapper} Klapper, I.; Hagstrom, R.; Fine, R; Sharp, K.; Honig,
B.  {\em PROTEINS:  Structure, Function, and Genetics} {\bf 1986}, 1,
47.

\bibitem{gilson_b} Gilson, M. K.; Sharp, K. A.;  Honig, B. H.
{\em J.  Comp.  Chem.} {\bf 1987}, 9, 327.

\bibitem{pascual-ahuir} Pascual-Ahuir, J.  L.; Silla, E.; Tomasi, J.;
Bonaccorsi, R.  {\em J.  Comp.  Chem.}  {\bf 1987}, 8, 778.


\bibitem{rashin_a} Rashin, A.  A.;  Namboodiri, K.  {\em J.  Phys.
Chem.}, {\bf 1987}, 91, 6003.


\bibitem{gilson_c} Gilson, M. K.;  Honig, B. H. {\em PROTEINS:
Structure, Function, and Genetics} {\bf 1988}, 3, 32.

\bibitem{zauhar_b} Zauhar, R.  J.;  Morgan, R.  S.  {\em J.  Comp.
Chem.}  {\bf 1988}, 9, 171.


\bibitem{davis} Davis, M.  E.; McCammon, J.  A.  {\em J.  Comp.  Chem.}
{\bf 1989}, 10, 386.

\bibitem{yoon-lenhoff} Yoon, B.  J.;  Lenhoff, A.  M.  {\em J.  Comp.
Chem.}  {\bf 1990}, 11, 1080.


\bibitem{juffer} Juffer, A.  H.; Botta, E.  F.  F.; van Keulen, B.  A.
M.; van der Ploeg, A.;  Berendsen, H.  J.  C.  {\em J.  Comp.  Phys.}
{\bf 1991}, 97, 144.

\bibitem{nicholls} Nicholls, A.; Honig, B. {\em J.  Comp.  Chem.}
{\bf 1991}, 12, 435.

\bibitem{snitzer} Snitzer, J.  E.; Lambrakis, K.  C.  {\em J.  Theor.
Biol.}  {\bf 1991}, 152, 203.





\bibitem{wang} Wang, B.; Ford, G. P. {\em J.  Chem.  Phys.} {\bf
1992}, 97, 4162.


\bibitem{oberoi} Oberoi, H.; Allewell, N.  M.  {\em Biophys.  J.}  {\bf
1993}, 65, 48.

\bibitem{zhou}   Zhou, H.-X. {\em Biophys.  J.} {\bf 1993}, 65, 955.


\bibitem{mdmp} Mohan, V.; Davis, M.  E.; McCammon, J.  A.;  Pettitt,
B.  M.  {\em J.  Phys.  Chem.}  {\bf 1992}, 96, 6428.

\bibitem{you} You, T.  J.;  Harvey, S.  C.  {\it J.  Comp.  Chem.}
{\bf 1993}, 14, 484.


\bibitem{simonson} Simonson, T.; Br\"unger, A.  T.  {\em J.  Phys.
Chem.}  {\bf 1994}, 98, 4683-4694.

\bibitem{tucker} Tucker, S.  C.; Gibbons, E.  M.  {\em Structure and
reactivity in aqueous solution:  Characterization of chemical and
biological systems,} edited by C.  J.  Cramer and D.  G.  Truhlar, {\em
ACS Symposium Series} {\bf 568}, 198,(1994).  (ACS, Washington DC,
1994).

\bibitem{bharadwaj} Bharadway, R.; Windemuth, A.; Sridharan, S.; Honig,
B.; Nicholls, A.; {em J.  Comp.  Chem.}  {\bf 1995}, 16, 898.


\bibitem{phg} Pratt, L. R.; Hummer, G.;  Garc\'{\i}a, A.  E. {\em
Biophys.  Chem.} {\bf 1994}, 51, 147.

\bibitem{tp} Tawa, G.~J.; Pratt, L.~R.  In Cramer, C.~J.; Truhlar,
D.~G., Eds.  {\em Structure and Reactivity in Aqueous Solution:
Characterization of Chemical and Biological Systems.  {ACS} Symposium
Series}; ACS:  Washington DC, 1994, Vol.~568, p.\ 60.

\bibitem{lbk} Levy, R.  M.; Belhadj, M.;  Kitchen, D.  B.  {\em J.
Chem.  Phys.}  {\bf 1991}, 95, 3627.

\bibitem{wilfred} Smith, P.  E.; .  van Gunsteren, W.  F {\em J.  Chem.
Phys.}  {\bf 1994}, 100, 577.


\bibitem{hrl} Hirata, F.; Redfern, P.; Levy, R.  M.  {\it Int.  J.
Quant.  Chem.\/} {\bf 1988}, 15, 179.

\bibitem{fbl} Figueirido, F.; Del Bueno, G.  S.; Levy, R.  M.  {\it
Biophys.  Chem.\/} {\bf 1994}, 51, 235.


\bibitem{kw} Tawa, G.~J.;  Pratt, L. R. {\em J.  Am.  Chem.
Soc.} {\bf 1995}, 117, 1625.

\bibitem{hpga} Hummer, G.; Pratt, L. R.;  Garc\'{\i}a, A.  E.
LA-UR-95-1161, {\em J.  Phys.  Chem.} in press (1995).
chem-ph\# 9505005.

\bibitem{hpgb} Hummer, G.; Pratt, L. R.;  Garc\'{\i}a, A.  E.
LA-UR-95-1612, {\em J.  Phys.  Chem. } in press (1995).
chem-ph\# 9507004.

\bibitem{hammersley} Hammersley, J.  M.; Handscomb, D.  C.  {\it Monte
Carlo Methods\/}, Chapman and Hall, London, 1964, pp.  31-36.


\bibitem{keng} Keng, H.  L.; Yuan, W.  APPLICATIONS OF NUMBER THEORY TO
NUMERICAL ANALYSIS, (Springer-Verlag, NY, 1981).

\bibitem{lubotzky} Lubotzky, A.; Phillips, R.; Sarnak, P.  {\it Comm.
Pure Appl.  Math.\/} {\bf 1986}, 39, S149; {\it Comm.  Pure Appl.
Math.\/} {\bf 1987}, 40, 401.

\bibitem{tichy} Tichy, R.  F.  {\it J.  Comp.  Appl.  Math.\/} {\bf
1990}, 31, 191.

\bibitem{niederreiter} Niederreiter, H.  {\it Random Number Generation
and Quasi-Monte Carlo Methods\/}, (SIAM, Philadelphia, 1992).

\bibitem{press} Press, W.  H.; Teukolsky, S.  A.; Vetterling, W.  T.;
Flannery, B.  P.  {\it Numerical Recipes, The Art of Scientific
Computing\/}, 2nd edition, Cambridge University Press, NY, 1992, \S 7.7.

\bibitem{rowlinson-widom} Rowlinson, J.  S.; Widom, B.  {\it Molecular
Theory of Capillarity\/}, (Clarendon Press, Oxford, 1989), pp.  19-20.

\bibitem{lsb} Lebowitz, J.  L.; Stell, G.; Baer, S.  {\it J.  Math.
Phys.\/} {\bf 1965}, 6, 1282.

\bibitem{rashin89} Rashin, A. A.; {\em J.  Phys.  Chem.} {\bf 1989},
93, 4664.

\bibitem{ford} Ford, G.  P.; Wang, B. {\em J.  Am.  Chem.  Soc.}  {\bf
1992}, 114, 10563.

\bibitem{thg} Shang, H.  S.; Head-Gordon, T.  {\it J.  Am.  Chem.
Soc.\/} {\bf 1994}, 116, 1528.

\bibitem{connolly} Connolly, M.  L.  {\it J.  Appl.  Cryst.\/} {\bf
1985}, 18, 499.

\bibitem{jaswon} See, for example, Jaswon,  M.  A.  in {\it Topics in
Boundary Element Research\/}, edited by C.  A.  Brebbia, {\bf 1},
13(1984).

\bibitem{rashin85} Rashin, A. A.;  Honig, B. {\em J.  Phys.
Chem.} {\bf 1985}, 89, 5588.

\bibitem{lim} Lim, C.; Bashford, D.; Karplus M.  {\em J.  Phys.  Chem.}
{\bf 1991}, 95, 5610.


\bibitem{hummer92} Hummer, G; Soumpasis, D.  M.; Neumann, M.  {\bf
1994}, 81, 1155.

\bibitem{guardia91a} Gu\`{a}rdia, E.; Rey.  R.; Padr\'{o}, J.  A.  {\it
Chem.  Phys.\/} {\bf 1991}, 155, 187.

\bibitem{guardia91b} Gu\`{a}rdia, E.; Rey.  R.; Padr\'{o}, J.  A.  {\it
J.  Chem.  Phys.\/} {\bf 1991}, 95, 2823.

\bibitem{pr} Pettitt, B.  M.; Rossky, P.  J.  {\it J.  Chem.  Phys.\/}
{\bf 1986}, 84, 5836.

\bibitem{smith:94}   Smith, D.  E.;  Dang, L.  X.  {\it J.  Chem.
Phys.\/} {\bf 1994}, 100, 3757.

\bibitem{wljtb} Jorgensen, W.  L.; Buckner, J.  K.; Huston, S.  E.;
Rossky, P.  J.  {\it J.  Am.  Chem.  Soc.\/} {\bf 1987}, 109, 1891.

\bibitem{nadmpa} Huston, S.  E.; Rossky, P.  J.  {\it J.  Phys.
Chem.\/} {\bf 1989}, 93, 7888.

\bibitem{nadmpb} Chen, S.  W.; Rossky, P.  J.  {\it J.  Phys.  Chem.\/}
{\bf 1993}, 97, 6078.

\bibitem{nadmpc} Gorenstein, D.  G.; Findley, J.  B.; Luxon, B.  A.;
Kar, D.  {\it Biochim.  Biophys.  Acta\/} {\bf 1977}, 475, 184.

\bibitem{nadmpd} Friedman R.  A.; Mezei, M.  {\it J.  Chem.  Phys.\/}
{\bf 1995}, 102, 419.

\bibitem{amber} Weiner, S.  J.; Kollman, P.  A.; Case, D.  A.; Singh, U.
C.; Ghio, C.; Alagona, G.; Profeta Jr., S.; Weiner, P.  {\it J.  Am.
Chem.  Soc.\/} {\bf 1984}, 106, 765.

\bibitem{sn2a} Chandrasekhar, J.; Smith, S.  F.; Jorgensen, W.  L.  {\it
J.  Am.  Chem.  Soc.\/} {\bf 1985}, 107, 154.

\bibitem{sn2b} Chiles, R.  A.; Rossky, P.  J.  {\it J.  Am.  Chem.
Soc.\/} {\bf 1984}, 106, 6867.

\bibitem{sn2c} Huston, S.  E.; Rossky, P.  J.; Zichi, D.  A.  {\it J.
Am.  Chem.  Soc.\/} {\bf 1989}, 111, 5680.

\bibitem{forma} Madura, J.  D.; Jorgensen, W.  L.  {\it J.  Am.  Chem.
Soc.\/} {\bf 1986}, 108, 2517.

\bibitem{formb} Yu, H.-A.; Karplus, M.  {\it J.  Am.  Chem.  Soc.\/}
{\bf 1990}, 112, 5706.

\bibitem{nmaaa} Jorgensen, W.  L.; Gao, J.  {\it J.  Am.  Chem.  Soc.\/}
{\bf 1988}, 110, 4212.

\bibitem{nmaab} Radom, L.; Riggs, N.  V.  {\it Aust.  J.  chem.\/} {\bf
1982}, 35, 1071.

\bibitem{nmaac} Yu, H.-A.; Pettitt, B.  M.; Karplus, M.  {\it J.  Am.
Chem.  Soc.\/} {\bf 1991}, 113, 2425.

\bibitem{sharp} Sharp, K.; A.  Jean-Charles,  Honig, B. {\em J.
Phys.  Chem.} {\bf 1992}, 96, 3822.

\bibitem{NR} Press, W.  H.; Teukolsky, S.  A.; Vetterling, W.  T.;
Flannery, B.  P.  {\it Numerical Recipes, The Art of Scientific
Computing\/}, 2nd edition, Cambridge University Press, NY, 1992, \S 2.6.

\bibitem{estruc} (a) Klopman, G.  {\em Chem.  Phys.  Lett.  } {\bf
1967}, 1, 200.  (b) Rinaldi, D.; Rivail, J.  L.  {\em Theor.  Chim.
Acta } {\bf 1973}, 99, 4899; (c) Germer, H.  A.  {\em Theor.  Chim.
Acta } {\bf 1974}, 34, 145; (d) Hylton, J.; Christoffersen, R.  E.;
Hall, G.  C.  {\em Chem.  Phys.  Lett.  } {\bf 1974}, 26, 501; (
Miertus, S.; Kysel, O.  {\em Chem.  Phys.  } {\bf 1977}, 21, 27; (f) S.
Miertus, S.; Scrocco, E.;  Tomasi, J.  {\em Chem.  Phys.  } {\bf
1981}, 55, 117; (g) Bonaccorsi, R.; Cimiraglia, R.; Tomasi, J.  {\em J.
Comput.  Chem.  } {\bf 1983}, 4, 567; (h) Russel, S.  T.; Warshel, A.
{\em J.  Mol.  Biol.  } {\bf 1985}, 185, 389; (i) Hoshi, H.; Sakurai,
M.; Inoue, Y.; Chujo, R.  {\em J.  Chem.  Phys.  } {\bf 1987}, 87, 1107;
(j) Mikkelsen, K.; Dalgaard, E.; Swanstrom, J.  {\em J.  Phys.  Chem.  }
{\bf 1987}, 91, 3081; (k) Aguilar, M.  A.; Olivares delValle, F.; {\em
J.  Chem.  Phys.  } {\bf 1989}, 129, 439; (l) King, G.; Warshel, A.;
{\em J.  Chem.  Phys.  } {\bf 1989}, 91, 3647; (m) Chudinov, G.  E.;
Napolov, D.  V.; Basilevsky, M.  V.  {\em Chem.  Phys.  } {\bf 1991},
160, 41; (n) Agren, H.; Mikkelsen, K.  V.  {\em J.  Mol.  Struct.
(Theochem) } {\bf 1991}, 234, 425; (o) Wong, M.  W.; Frisch, M.  J.  ;
Wiberg, K.  B.  {\em J.  Am.  Chem.  Soc.  } {\bf 1991}, 113, 4776; (p)
Cramer, C.  J.  ; Truhlar, D.  G.  {\em J.  Am.  Chem.  Soc.  } {\bf
1991}, 113, 8305; (q) Ford, G.  P.  ; Wang, B.  {\em J.  Comput.  Chem.
} {\bf 1992}, 13, 229.  (r) Rauhut, G.; Clark, T.; Steinke, T.  {\em J.
Am.  Chem.  Soc.  } {\bf 1993}, 115, 9174; (s) Rosch, N.; Zerner, M.  C.
{\em J.  Phys.  Chem.  } {\bf 1994}, 98, 5817; (t) Cramer, C.  J.;
Truhlar, D.  G.  {\em J.  Comp-aid Molec.  Des.  } {\bf 1992}, 6, 629;
(u) Rashin, A.  A.; Young, L.; Topol,I.  A.  {\em J.  Biophys.  Chem.  }
{\bf 1994}, 51, 359; (v) Rashin, A.  A.; Bukatin, M.  A.; Andzelm, J.
N.; Hagler, A.  T.  {\em J.  Biophys.  Chem.  } {\bf 1994}, 51, 375; (w)
Tomasi, J.; Bonaccorsi, R.; Cammi, R.; Olivares del Valle, F.  {\em J.
Mol.  Struct.  (Theochem) } {\bf 1991}, 234, 401; (x) Tapia, O.  {\em
Theor.  Chim.  Acta } {\bf 1978}, 47, 157; (y) Tapia, O.; Poulain,E.;
Sussman, F.  {\em Theor.  Chim.  Acta } {\bf 1978}, 47, 171; (z)
Mikkelsen, K.  V.; Agren, H.; H.  Hensen, J.  A.; Helgaker, T.  U.  {\em
J.  Chem.  Phys.  } {\bf 1988}, 89, 3086; (aa) Rivail, J.  L.  {\em
Compt.  Rend.  Acad.  Sci.  (Paris) } {\bf 19990}, 311, 307; (bb)
Olivares del Valle, F.  J.; Tomasi, J.  {\em J.  Chem.  Phys.  } {\bf
1991}, 150, 139; (cc) Aguilar, M.  A.; Olivares del Valle, F.  J.;
Tomasi, J.  {\em J.  Chem.  Phys.  } {\bf 1991}, 150, 151; (dd) Olivares
del Valle, F.  J.; Bonaccorsi, R.; Cammi, R.;  Tomasi, J.  {\em J.
Mol.  Struct.  (Theochem) } {\bf 1991}, 230, 295.  (ee) Frisch, M.  J.;
Head-Gordon, M.; Trucks, G.  W.; Foresman, J.  B.; Schlegel, H.  B.;
Raghavachari, K.  ; Robb, M.  A.; Wong, M.  W.; Binkley, S.  J.; Seeger,
R.  ; Melius, R.  C.  F.; Baker, J.; Martin, R.  L.; Kahn, L.  R.;
Stewart, J.  J.  P.; Topiol, S.; Pople, J.  A.  GAUSSIAN 91 (1991); (ff)
Wong, M.  W.; Wiberg, K.  B.; Frisch, M.  J.  {\em J.  Am.  Chem.  Soc.
} {\bf 1992}, 114, 523; (gg) Wong, M.  W.; Wiberg, K.  B.; Frisch, M.
J.  {\em J.  Am.  Chem.  Soc.  } {\bf 1992}, 114, 1645; (hh) Olivares,
F.  J.; Aguilar, M.  A.  {\em J.  Comput.  Chem.  } {\bf 1992}, 13, 115;
(ii) Chipot, C.; Rinaldi, D.; Rivail, J.  L.  {\em Chem.  Phys.  Lett.
} {\bf 1992}, 191, 287; (jj) Chipot, C.; Gorb, L.  G.; Rivail, J.  L.
{\em J.  Phys.  Chem.  } {\bf 1994}, 98, 1601; (kk) Chen, J.  L.;
Noodleman, L.; Case, D.  A.; Bashford, D.  {\em J.  Phys.  Chem.  } {\bf
1994}, 98, 11059; (ll) Tawa, G.~J.; Martin, R.  L.; Pratt, L.  R.;
Russo, R.  V.  LA-UR-94-3425, submitted {\em J.  Phys.  Chem.  } 1995;
(mm) Tannor, D.  J.; Marten, B.; Murphy, R.; Friesner, R.  A.; Sitkoff,
D.; Nicholls, A.; Ringhalda, M.; Goddard, W.A.; Honig, B.  {\em J.  Am.
Chem.  Soc.  } {\bf 1994}, 116, 11875.

\bibitem{yu-karplus} Yu, H.-A.; Karplus, M.  {\it J.  Chem.  Phys.\/}
{\bf 1988}, 89, 2366.

\bibitem{king} King, G.; Barford, R.  A.  {\it J.  Phys.  Chem.\/} {\bf
1993}, 97, 8798.

\bibitem{bagchi} Chandra, A.; Bagchi, B.  {\it J.  Chem.  Phys.\/} {\bf
1989}, 91, 3056.

\bibitem{kirzhnits} Dolgov, O.  V.; Kirzhnits, D.  A.; Maksimov, E.  G.
{\it Rev.  Mod.  Phys.\/} {\bf 1981}, 53, 81.

\bibitem{green2} Bronshtein, I.  N.; Semendyayev, K.  A.  {\it
HANDBOOK OF MATHEMATICS\/} (Van Nostrand Reinhold, NY, 1985), p.  460.

\bibitem{stereo} Bronshtein, I.  N.; Semendyayev, K.  A.  {\it HANDBOOK
OF MATHEMATICS\/} (Van Nostrand Reinhold, NY, 1985), p.  476.

\bibitem{brush} Brush, S.  G.; Sahlin, H.  L.; Teller, E.  {\it J.
Chem.  Phys.\/} {\bf 1966}, 45, 2102.

\bibitem{nijboer} Nijboer, B.  R.  A.; Ruijgrok, T.  W.  {\it J.  Stat.
Phys.\/} {\bf 1988}, 53, 361.

\end{references}
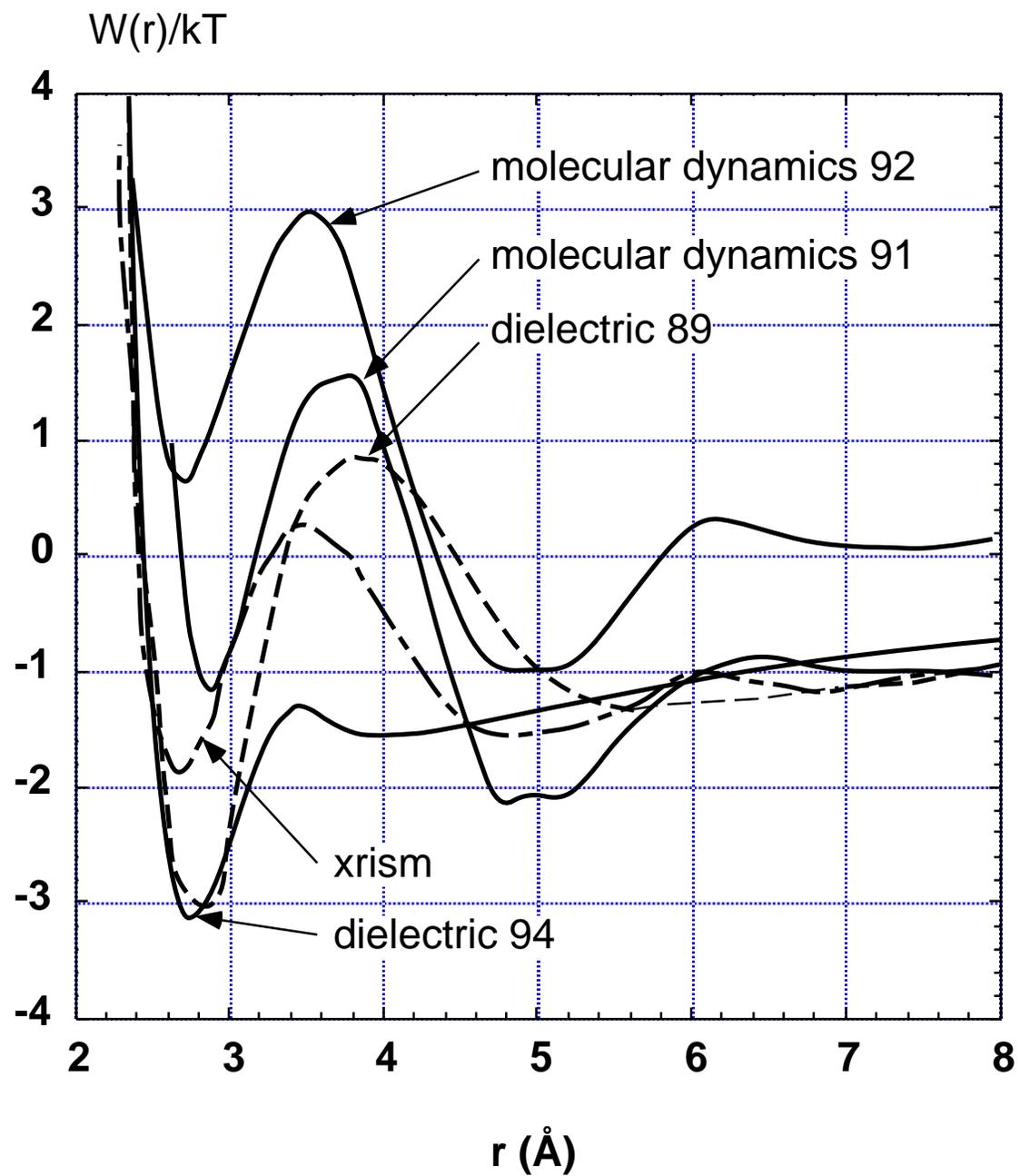
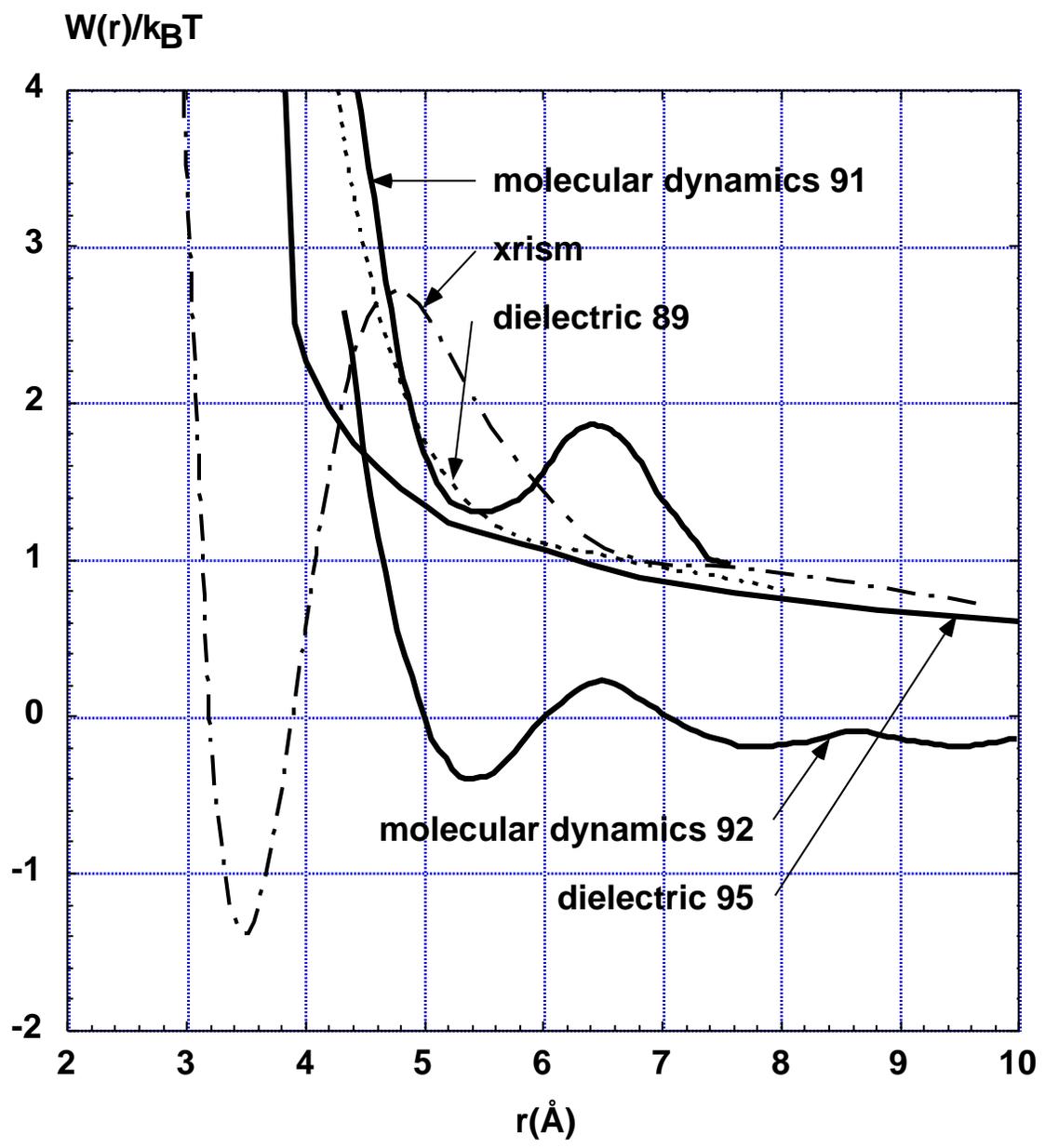
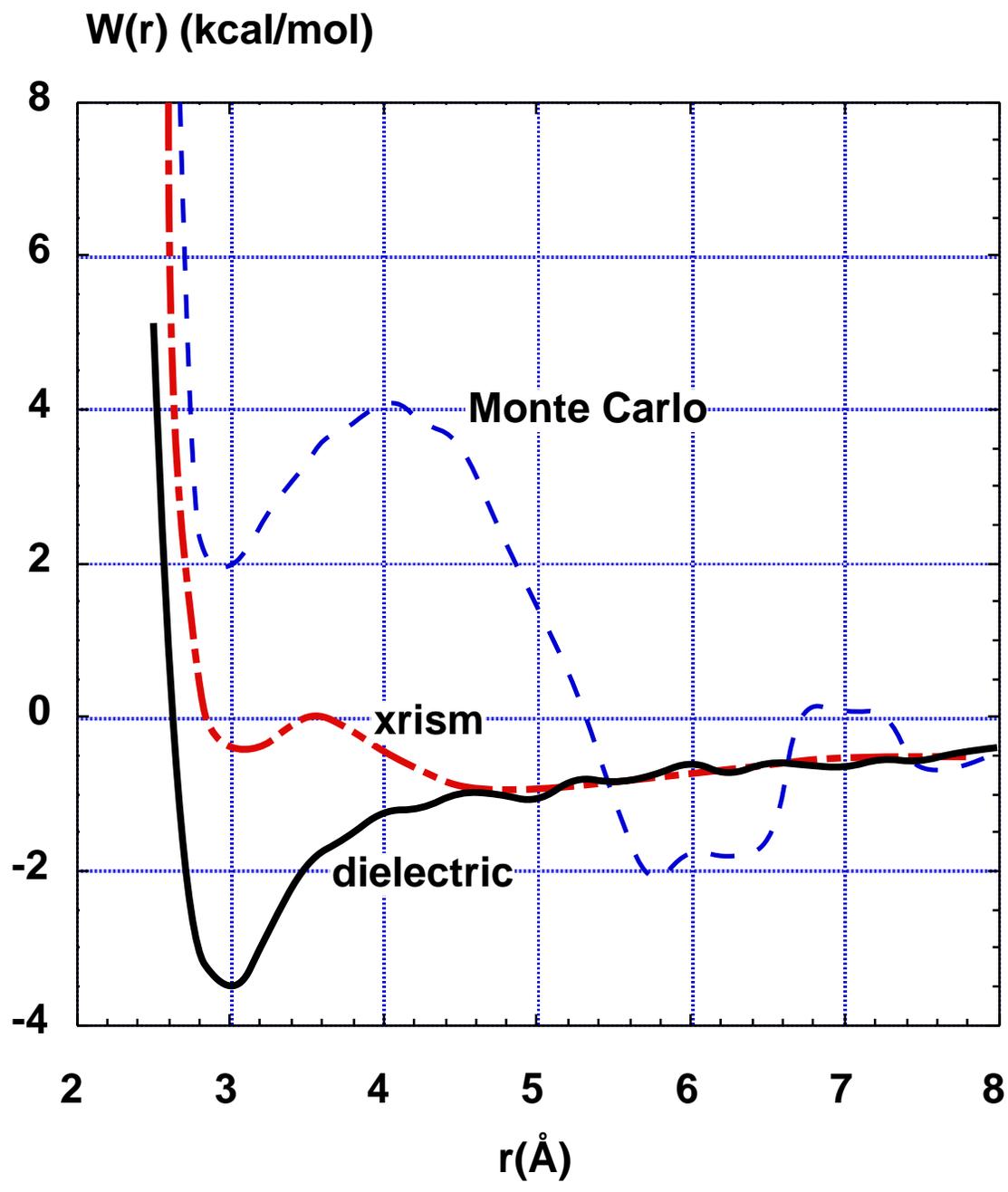
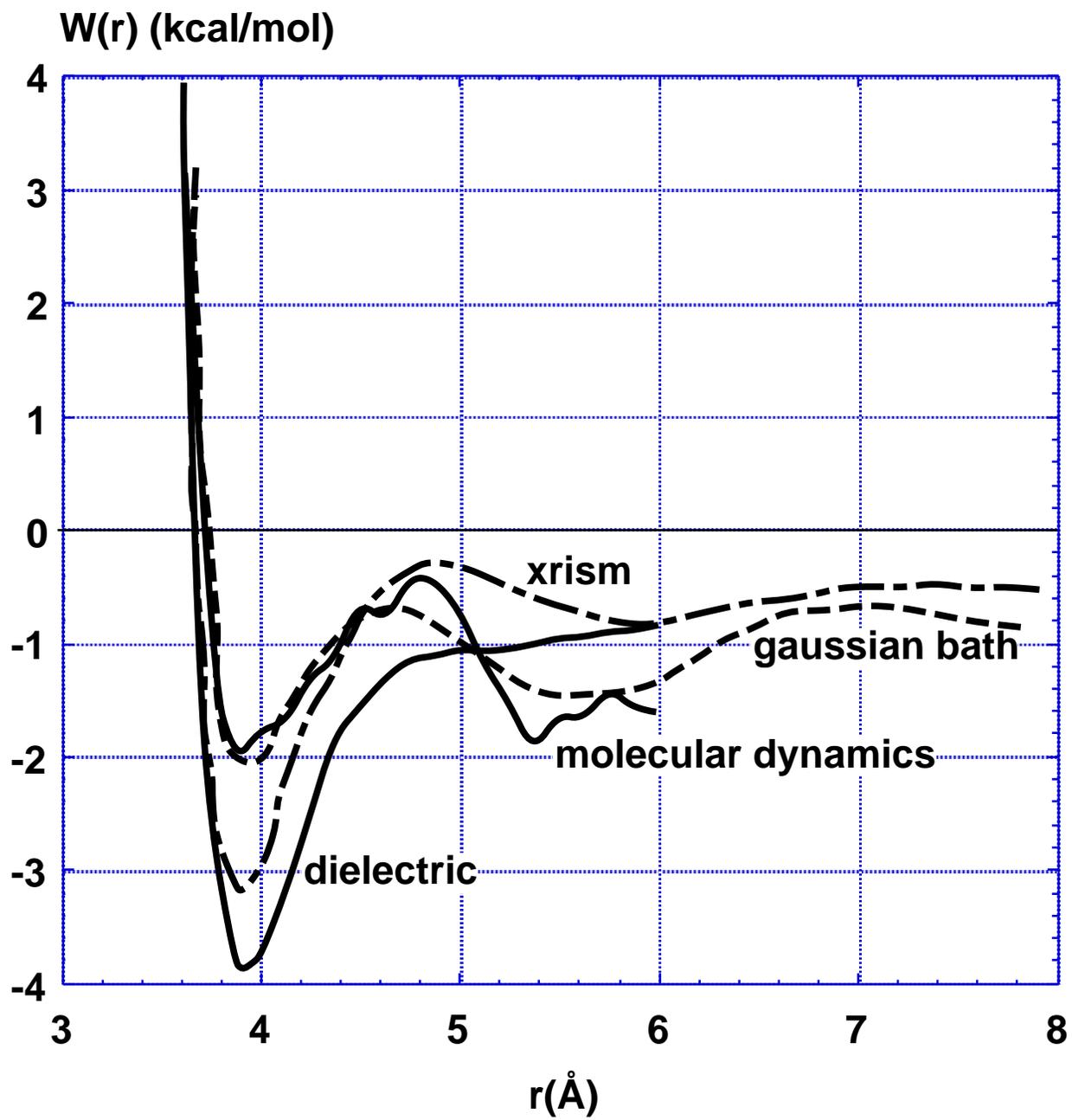
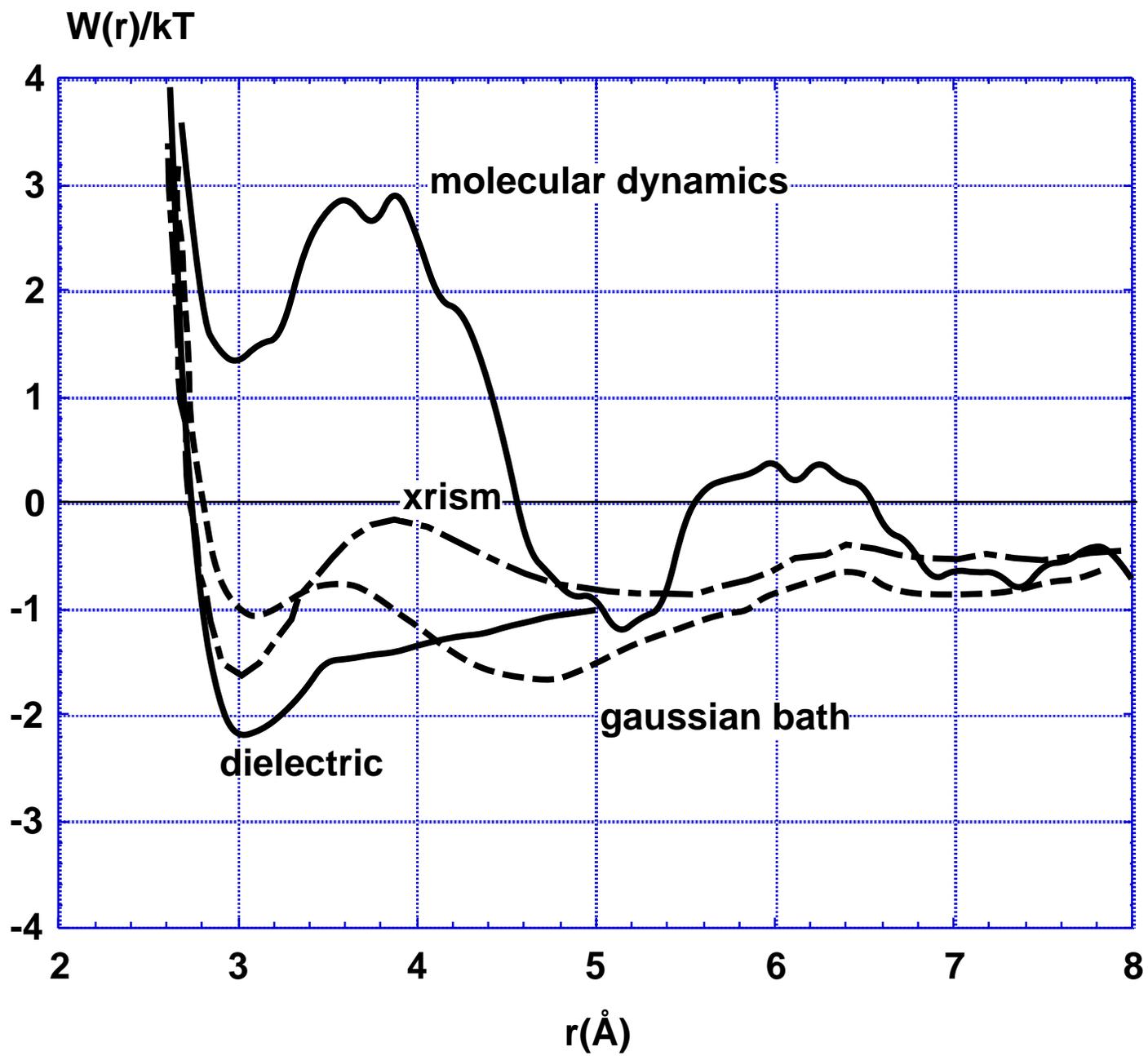
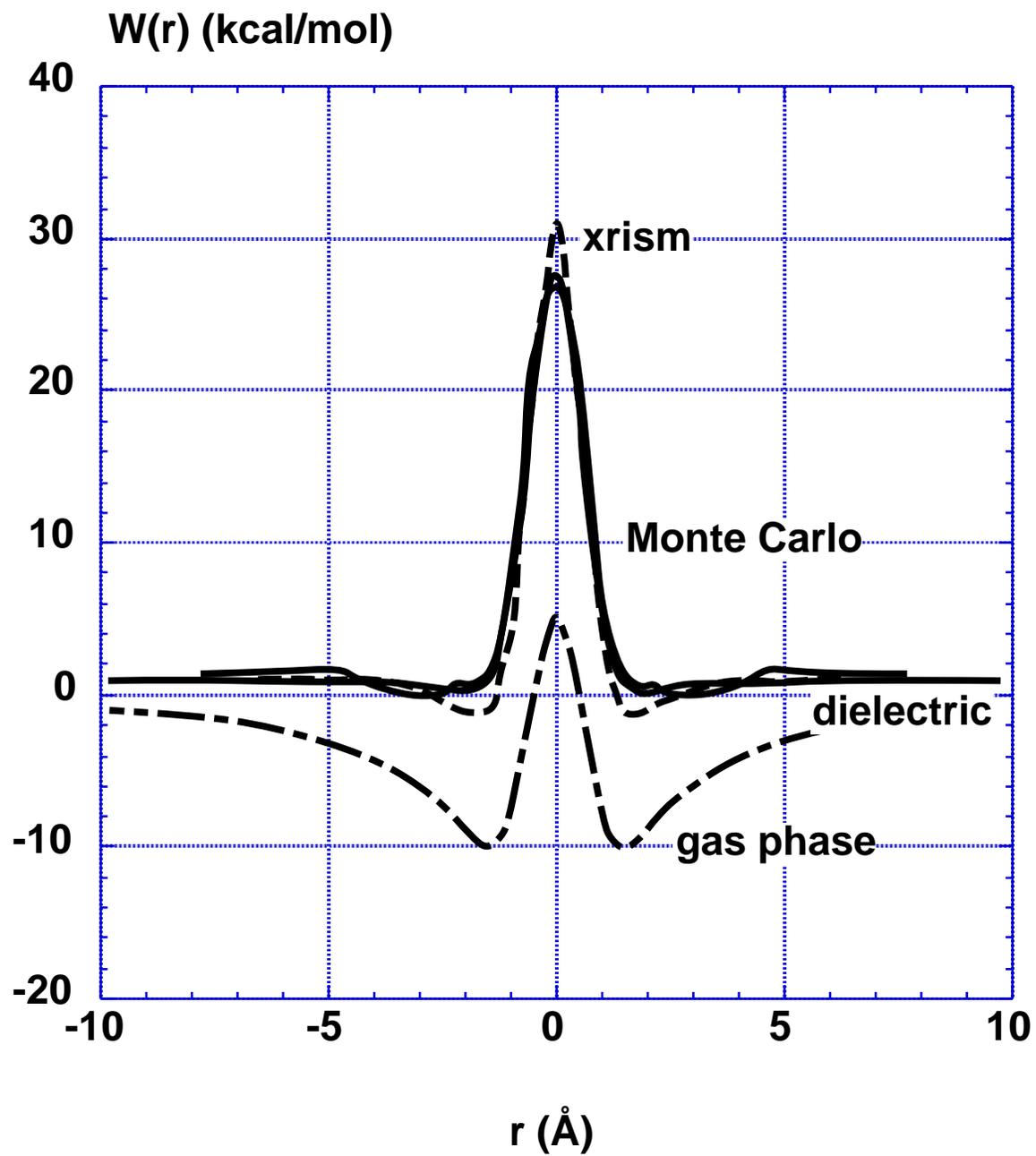
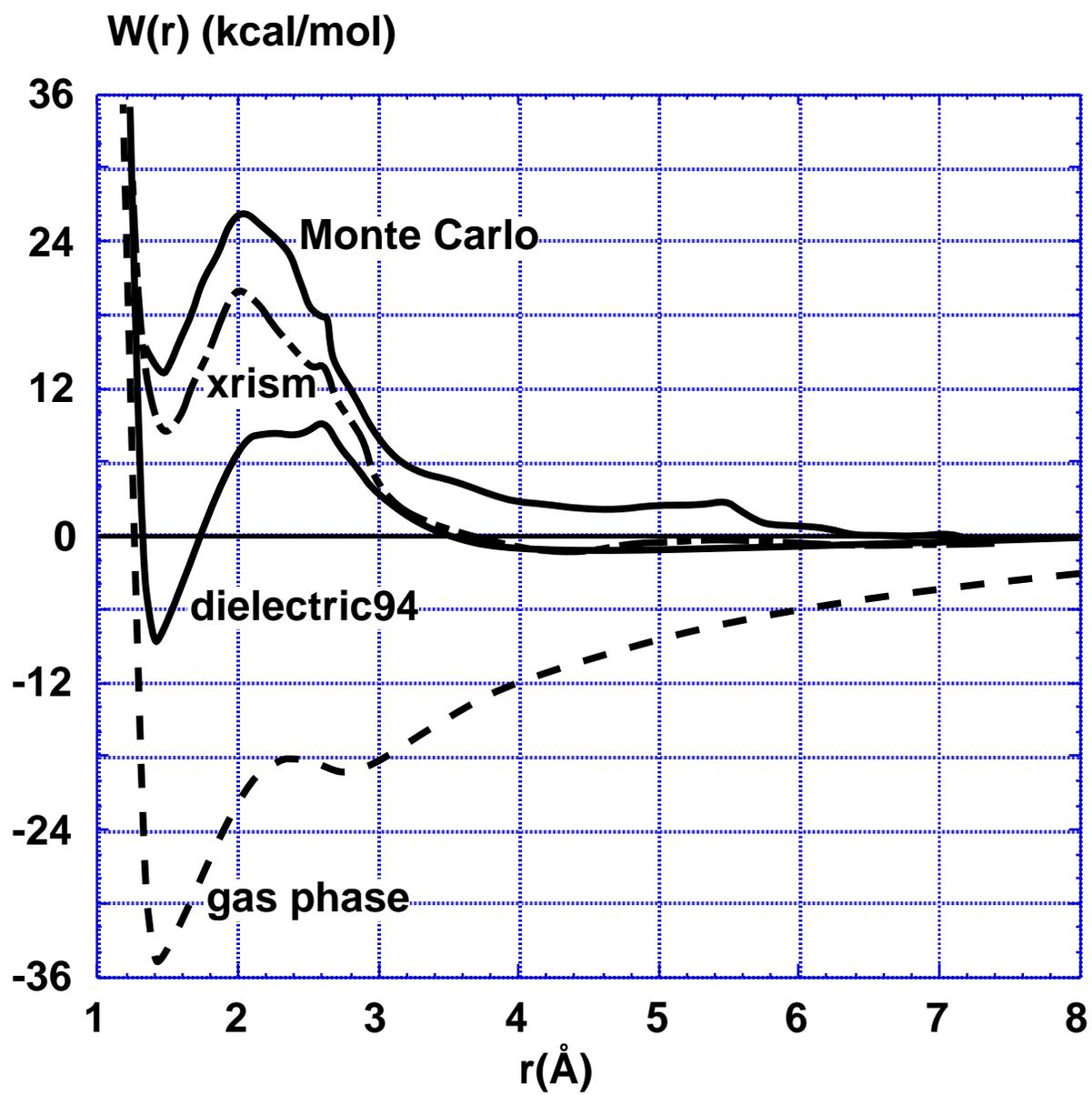
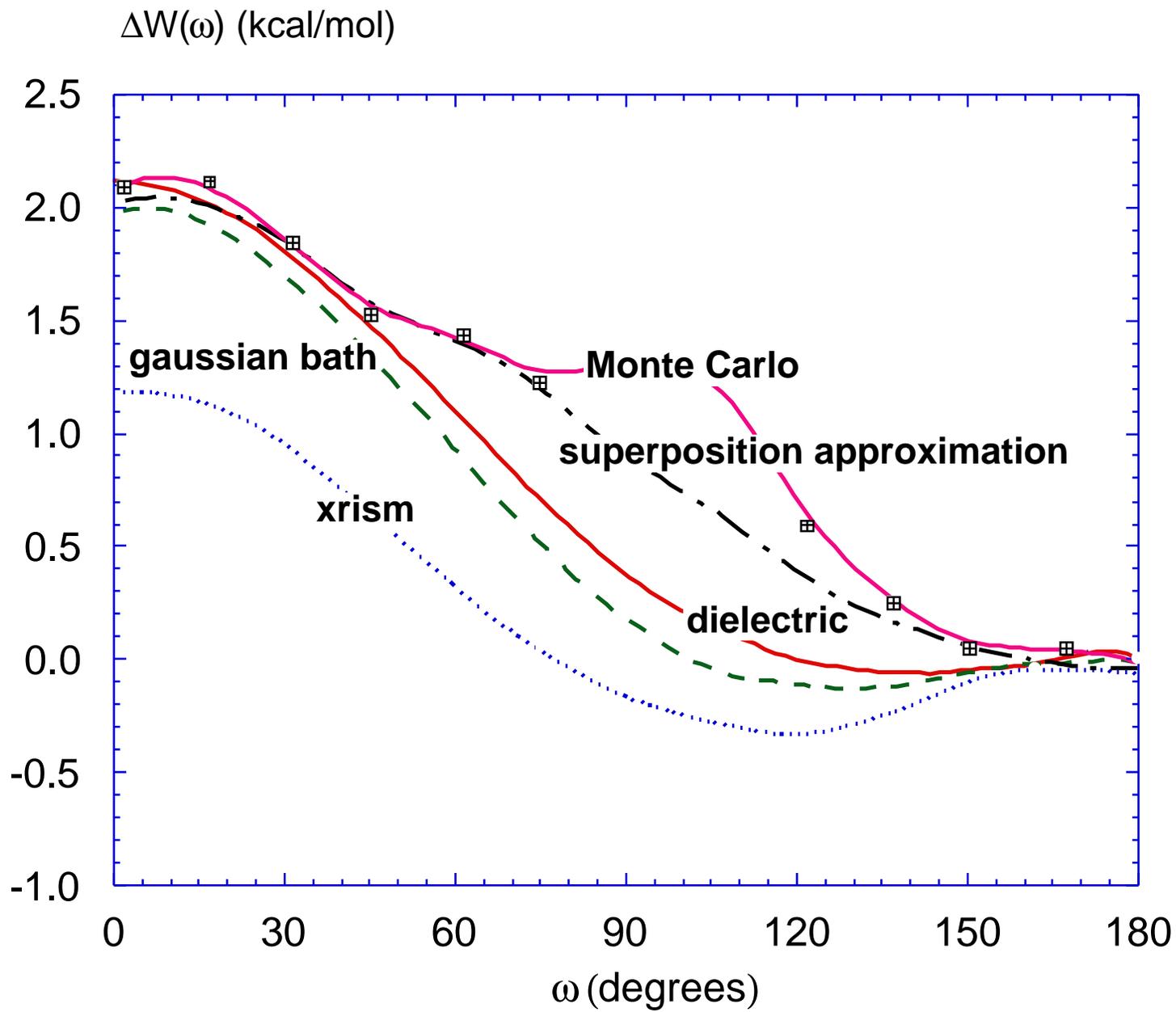
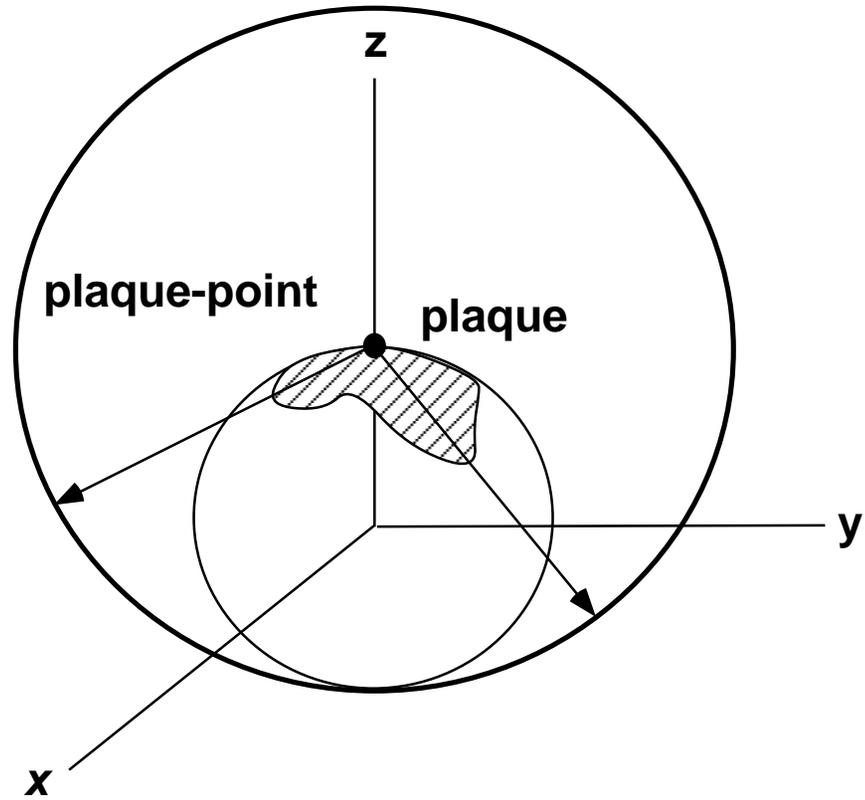
\end{document}